\documentclass[aps,prb,superscriptaddress,10pt,twocolumn]{revtex4-2}
\usepackage{color}
\usepackage{amsmath} 
\usepackage{amssymb}
\usepackage{amsbsy}
\usepackage{bbold}
\usepackage{bm}
\usepackage{ctable}
\usepackage[ansinew]{inputenc} 

\renewcommand{\vec}[1]{\boldsymbol{#1}}

\usepackage{graphicx}

\begin{document}
\title{Superconductor-``Metal" Transition of One-dimensional \\
Interacting Bosons
with Ohmic Quantum Dissipation}
\author{Miguel A. Cazalilla}
\affiliation{Donostia International Physics Center (DIPC), 20018 Donostia-San Sebastian, Spain}
\affiliation{Ikerbasque, Basque Foundation for Science, 48013 Bilbao, Spain}

\begin{abstract}
The phase diagram of a system of interacting bosons (Cooper pairs)  hopping on a one-dimensional (1D) lattice with onsite phase dissipation describing the tunneling to a nearby diffusive normal-metal electrode is studied. Starting from  the system at commensurate lattice filling, it is shown by a combination of analytical techniques that the phase diagram contains  two quantum phases: A dissipative Bose-Einstein condensate (D-BEC) or superconductor with long-range phase coherence, and a dissipative Mott insulator (D-Mott) or "metal"  with exponentially decaying phase correlations in space and local imaginary-time correlations decaying as the local pairing correlations of the electrode. The D-Mott/``metal" phase can be  described as a 1D array of dissipative boson puddles, weakly coupled by Josephson tunneling. The puddle size roughly corresponds to the length scale beyond which  phase slips suppress phase coherence. The dissipative time-dependent Ginsburg-Landau  theory phenomenologically used by Sachdev, Werner, and Troyer  [Phys. Rev. Lett. {\bf 92} 237003 (2004)] for the superconductor-metal transition in quasi-1D wires is derived from this microscopic puddle picture. Thus, the criticality of the D-Mott/D-BEC transition is shown to belong to the Wilson-Fisher universality class with dynamical exponent $z\approx 2$.  At small doping, the D-Mott/metal phase remains stable due to its finite compressibility, which is computed to leading order  in a perturbation expansion of the dissipation strength and the inter-puddle Josephson coupling. At larger doping,  using a mapping to a pseudospin chain combined with bosonization,  the D-BEC/superconductor phase is the ground state for non-vanishing but arbitrarily small dissipation.   Similarities and differences with deconfinement transition  of an array 1D bosonic Mott insulators in  anisotropic optical lattices are also discussed. 
\end{abstract}
\maketitle

\section{Introduction}

Superconductor-metal transitions in (quasi) one-dimensional (1D) systems have attracted much theoretical attention  over the years~\cite{FEIGELMAN1998107,Larkin_PhysRevLett.86.1869,Sachdev_PhysRevLett.92.237003,Werner_JPSP,Lobos1_PhysRevB.80.214515,Lobos3_PhysRevB.84.024523,Refael_PhysRevLett.101.106402}.
Assuming the validity of  time-dependent Ginsburg-Landau
 (TDGL) theory~\cite{Tucker_PhysRevB.3.3768}  all the way to zero temperature,  Sachdev \emph{et al.}~\cite{Sachdev_PhysRevLett.92.237003} used a quantized version of a dissipative TDGL  as a minimal description for the superconductor-metal quantum phase transition in quasi-1D wires. In their theory, quantum dissipation is 
phenomenologically introduced to describe the decay of Cooper pairs into normal electrons  in multi-channel wires with a non-uniform pairing potential. The  transition was found~\cite{Sachdev_PhysRevLett.92.237003,Werner_JPSP} to be in the O$(2)$  Wilson-Fisher universality class~\cite{Pankov_PhysRevB.69.054426} with dynamical exponent $z\approx 2$.
These authors also emphasized the
 differences of this superconductor-metal transition with the superconductor-(Mott) insulator transition in 1D~\cite{Giamarchi_book},
 which is driven by topological excitations (phase slips).
  
A different theory of the transition was developed  in Ref.~\cite{Lobos1_PhysRevB.80.214515} by Lobos and coworkers
and applied to a quasi-1D  superconducting wire with a tunneling coupling to a 2D diffusive metal. Leaking
of Cooper pairs in and out of the 1D wire leads to quantum dissipation~\cite{Cazalilla_PhysRevLett.97.076401}.
In the regime in which phase dissipation is local~\cite{FEIGELMAN1998107, Larkin_PhysRevLett.86.1869,Cazalilla_PhysRevLett.97.076401,Lobos1_PhysRevB.80.214515}, the 2D metal  behaves as an ohmic  bath that couples to the wire phase fluctuations. The authors of Ref.~\cite{Lobos1_PhysRevB.80.214515} used  bosonization~\cite{GogolinTsvelik1999,Giamarchi_book,Cazalilla_JphysB}, which describes
phase slips using sine-Gordon terms in the effective low-energy 
action~\cite{Giamarchi_book,Cazalilla_RevModPhys.83.1405}. Their competition with the local 
dissipation can be studied  using  perturbative renormalization-group~\cite{Lobos1_PhysRevB.80.214515}. This theory makes  explicit  
the  role  played by phase slips. However, it cannot be used to determine whether the transition between the two phases  is discontinuous or continuous and, in the latter case, which  universality class it belongs to. 

The lack of a clear  connection between the two different theories  described above is rather  unsatisfactory. We are confronted with two different choices: 1) An \emph{ad hoc}  quantized dissipative
TDGL theory that predicts a continuous transition in the Wilson-Fisher universality class~\cite{Sachdev_PhysRevLett.92.237003,Werner_JPSP} but  whose largely phenomenological nature obscures  its applicability range  as well as the microscopic origin of  the  dissipation, or 2) a bosonized field theory that accounts for some of the microscopic mechanisms for both  quantum dissipation and the destruction superconductivity by phase slips, but it is unable to predict the properties of the superconductor-metal quantum phase transition. The main goal of this work is to clarify the relationship between  these two theories  as well as  with earlier work~\cite{FEIGELMAN1998107,Larkin_PhysRevLett.86.1869}. Specifically, 
we will  show that the quantized dissipative TDGL theory applies to 1D systems at zero temperature, provided the distinction between the 1D system and  dissipative bath is made sufficiently clear and the properties of the latter are well specified. 

Indeed, in a broader context,   1D systems coupled to  dissipative baths must be regarded as open quantum many-body systems. The study of the latter has attracted a great deal of attention
in recent years, especially after the realization that quantum dissipation can be a
useful resource  to engineer exotic quantum states in quantum simulators of ultracold atomic gases~\cite{SyassenDuerr2008,Huang_PhysRevResearch.5.043192,TomitaTakahashi2017,Uchino2022,Dupuis2026}. Before the recent explosion of interest in  open (mainly Markovian) many-body 1D systems,  two physical realizations of 1D systems  coupled to an ohmic bath were studied and shown to lead to interesting quantum states in Ref.~\cite{Cazalilla_PhysRevLett.97.076401}. One realization (henceforth referred to as model I) considered a  1D wire capacitively coupled to a nearby 2D or 3D metallic gate. It was
shown that the dynamic screening of the
wire electrons by the density fluctuations in the gate results in electron  backscattering. 
The latter effectively results in the coupling of an ohmic dissipative environment to the density fluctuations of the wire with momentum $q = 2 p k_F$ ($p$ being an integer and $k_F$  the Fermi momentum). The ohmic dissipation is provided by the critical density fluctuations in the metal gate, which is assumed to be a (clean or diffusive) Fermi liquid.  It was shown later that  this type of dissipative environment  can be also realized  by loading in  optical lattices of mixed dimensionality an  interacting Bose-Fermi mixture~\cite{Malatsetxebarria_PhysRevA.88.063630}. 

A different realization  of  a 1D system coupled to a(-n ohmic) dissipative environment (referred to  as model II below)  was also proposed and analyzed in  Ref.~\cite{Cazalilla_PhysRevLett.97.076401} and it is relevant to our study here. It considered 
a (fluctuating) 1D superconductor~\cite{Luther_PhysRevLett.33.589} from which Cooper pairs can leak in and out via Josephson tunneling to a nearby diffusive metal electrode. The Cooper pairs propagate as critical pairing fluctuations in the electrode and, in the regime where the coherence length of the 1D wire is longer than the mean-free path in the diffusive metal electrode, a local ohmic coupling is generated for the wire's phase fluctuations. 

In the bosonization formalism, the imaginary-time actions describing both model I and II  are related by a duality transformation~\cite{GogolinTsvelik1999,Giamarchi_book}
 where the phase and density fields are interchanged. Relying on  this duality, a general phase diagram for the  models I and II was  proposed in Ref.~\cite{Cazalilla_PhysRevLett.97.076401}.  By a combination of different techniques, 
 it was concluded that the models have three different  phases: A Tomonaga-Luttinger liquid  (TLL)
with zero-temperature power-law correlations, a phase with true 
long-range order at zero temperature (which is crystal for  model I
and a Bose-Einstein Condensate for  model II), and a disordered phase with
exponential spatial correlations but power-law $\sim \tau^{-2}$ local correlations
for the density in model I (phase, in model II). Using a weak-coupling renormalization group analysis, the phase transition between the TLL  and the ordered phase was found in the Berezinskii-Kosterlitz-Thouless (BKT) universality class with dynamical exponent $z\approx 1$. However, the phase transition  
between the ordered and disordered phase was found  to be in the Wilson-Fisher universality class~\cite{Pankov_PhysRevB.69.054426,Sachdev_PhysRevLett.92.237003,Werner_JPSP}
 with dynamical exponent $z\approx 2$.

 The theoretical analysis carried out in Ref.~\cite{Cazalilla_PhfysRevLett.97.076401} 
 could not determine the universality class of the phase transition between the TLL
 and the disordered phase, or even establish whether there is a direct transition
 between these two phases at all.   On the other hand, recent numerical results~\cite{Rosso1_PhysRevB.107.165113} obtained using Langevin
 dynamics for the  field theory describing model I
  found a  BKT-like quantum phase transition 
 between a TLL and an ordered, crystal-like phase and  no evidence of a third
(disordered) phase. The disordered phase was observed in Monte Carlo
simulations of the classical 2D XY lattice model with anisotropic long-range interactions reported in Ref.~\cite{Werner_JPSP}, which in turn did not find the TLL phase.

 Factoring out that the two different numerical calculations may be tailored to search for phase transitions  with specific values of the dynamical  exponent ($z\approx 2$ in Ref.~\cite{Werner_JPSP}  and
 $z\approx 1$ in Ref.~\cite{Rosso1_PhysRevB.107.165113}), 
the puzzling  situation calls for revisiting the analysis of Ref.~\cite{Cazalilla_PhysRevLett.97.076401}. This provides a second 
motivation for this work and leads to  the discussion  in the concluding section,
which is written in hindsight after carefully 
analyzing a 1D quantum many-particle model that we believe faithfully realizes the classical 2D XY lattice model with local dissipation studied in Ref.~\cite{Werner_JPSP}. Finally, the physical realization 
considered here also allows to draw interesting parallels with the physics
of ultracold atoms. As a function of the
dissipative coupling, the phase transition
in this model shows also some similarities  (but also important differences) with the deconfinement transition
of 1D bosonic Mott insulators in an anisotropic optical lattice studied in Refs.~\cite{Cazalilla_2006,Ho_PhysRevLett.92.130405}. In the  model considered below the dissipative coupling is related to the Josephson tunneling of bosons in and out of a dissipative bath. Thus, the physical and mathematical analogies with the physics discussed in Refs.~\cite{Ho_PhysRevLett.92.130405,Cazalilla_2006} are unavoidable.

 The rest of the article is organized as follows: In the following section we
  introduce a 1D microscopic model of  a system of interacting bosons with Josephson coupling to an array ohmic baths. This model is
  analyzed in Sec.~\ref{sec:bosonization} combining bosonization and the weak
 coupling renormalization group  (RG) approaches.  Solving the weak coupling RG equations allows us to estimate the 
 phase boundary between a dissipative Bose-Einstein condensate  (D-BEC)  or superconductor and 
 a dissipative Mott (D-Mott) or ``metal'' phase. In the latter case, the term ``metal'' is used in a clear abuse of language  to relate to earlier work~\cite{Sachdev_PhysRevLett.92.237003,Werner_JPSP,Lobos1_PhysRevB.80.214515,
Larkin_PhysRevLett.86.1869,FEIGELMAN1998107}, where mathematically similar models
 have been considered in connection to the superconductivity of quasi-1D systems.
 The properties of the D-BEC/superconductor phase are explored
 in Sec.~\ref{sec:dbec} using the self-consistent harmonic approximation (SCHA). 
 In the D-Mott/``metal'' phase, we argue in Sec.~\ref{sec:dmott} that
 the system  can be regarded as an 1D array of weakly Josephson-coupled 
 boson puddles with the same local phase correlations 
 as the dissipative bath. Taking the continuum limit of this 
 lattice model of  puddles, a $1+1$ non-linear sigma model
 is derived in Sec.~\ref{sec:crotor}. When studied in the large-$N$ limit, 
 this model reproduces the correlation properties of the two phases and
 provides an estimate of the critical exponents at the quantum phase transition.  
 In Sec.~\ref{sec:GL}, we  derive the TDGL theory
 used in Ref.~\cite{Sachdev_PhysRevLett.92.237003}
 from the  model of coupled  puddles.
 Finally, in Sec.~\ref{sec:doping}, we consider the 
 case of non-integer filling. Using perturbation theory and a mapping to a pseudospin chain combined with bosonization, we show that, upon doping the D-Mott/metal phase,
 the D-Mott remains stable at small doping and  dissipation strength. However,  at large doping the ground state o is the D-BEC/superconductor phase for non-vanishing but arbitrarily small dissipation. The article is written to be as self-contained as possible and therefore  many details of the  calculations are provided in the Appendices.

\section{Model}\label{sec:model}
 
 In the absence of coupling to the dissipative environment, the system
interacting  bosons  in a one-dimensional lattice that we shall study below is  described by   the following  Hamiltonian:
\begin{align}
H_S &= -t \sum_{l} \left(  b^{\dag}_{l+1} b_l + b^{\dag}_{l} b_{l+1}\right) \notag \\
&\qquad + U \sum_{l} \left( n_l - n_0 \right)^2 + \sum_{l\neq l^{\prime}} V_{ll^{\prime}} n_l n_{l^{\prime}}.\label{eq:ham}
\end{align}
Here $n_l = b^{\dag}_l b_l$ ($l=1,\ldots,L$) is the site-occupation operator, where  the boson operators $b_{l},b^{\dag}_{l}$ obey
$[ b_{l}, b^{\dag}_{l^{\prime}} ] = \delta_{l,l^{\prime}}$, and  commute otherwise; $n_0  = N_B/L = n_0(\mu)$ is the lattice filling, $\mu$ being the chemical potential; 
$U$ is the on-site Hubbard interaction and $V_{ll^{\prime}}$ is the long-range tail of the interaction. This system 
is coupled via (Josephson) tunneling to a bath with which  bosons can be exchanged. For simplicity, we assume
the following linear coupling: 
\begin{equation}
H_D = -t_B \sum_{l=1}^{L} \left( b^{\dag}_l \Delta_l + \Delta^{\dag}_l b_l  \right). 
\end{equation}
The properties of the bath are entirely  specified by the (imaginary) time-ordered correlation functions of 
$\Delta_l$ and $\Delta^{\dag}_l$, which are assumed to be Gaussian fields that 
obey  $\langle \Delta_l \rangle = \langle \Delta^{\dag}_l \rangle = 0$, and have the following two-point correlations: 
\begin{align}
\langle \mathcal{T} [ \Delta_l(\tau) \Delta^{\dag}_{l^{\prime}}(0) ] \rangle &=  F(\tau) \delta_{l,l^{\prime}},\notag \\ 
\langle \mathcal{T}[ \Delta_l(\tau) \Delta_{l^{\prime}}(0)]\rangle  &= 0,
\end{align}
where $F(\tau) =  \tau^2_c \sum_{\omega_m\neq 0} |\omega_m| e^{-i\omega_m \tau}/\beta$. Here $\omega_m$ are the bosonic Matsubara frequencies where $\omega_m = 2 \pi m/\beta$,   $m$ being an integer. Asymptotically, $F(\tau)\sim (\tau_c/\tau)^2$ for $\beta \to +\infty$ and $|\tau|\gg \tau_c$, being $\tau^{-1}_c \lesssim t$ is a high frequency cut-off. Note that the dissipative coupling is local in space. A  realization of this model can be a  1D superconducting wire~\cite{Cazalilla_PhysRevLett.97.076401,Lobos_PhysRevB.86.035455} or Josephson junction array~\cite{FEIGELMAN1998107} 
where  Cooper pairs can leak to and from
a nearby diffusive  normal metal electrode, i.e.  model II of local dissipation introduced in Ref.~\cite{Cazalilla_PhysRevLett.97.076401} (see also~\cite{FEIGELMAN1998107,Larkin_PhysRevLett.86.1869} for arrays of superconducting islands). Throughout it will be assumed that the coupling is in the tunneling regime so that we can safely neglect the complications of the proximity effect in the electrode. As shown in Appendix~\ref{app:largenc}, this term can also be obtained in the limit where the boson fields are coupled to the pairing fluctuations of a large number of (independent) non-interacting fermion channels.

 Using the  correlators for $\Delta_l, \Delta^{\dag}_l$, the bath can be integrated 
out exactly within the imaginary-time path integral formalism. The latter allows us to write the partition function for the 
boson plus environment system as the following functional integral: 
\begin{equation}
Z = \int  \prod_l \left[d \bar{b}_l  d b_l \right] \, e^{-S\left[ b_l, \bar{b}_l \right] }.
\end{equation}
In units where the reduced Planck's constant equals unity, i.e. $\hbar = 1$,  the action $S\left[b_l, \bar{b}_l \right]$  is given by
\begin{align}
S\left[b_l, \bar{b}_l \right] &=  \int d\tau \sum_{l} \left[ \bar{b}_l \partial_{\tau} b_l - t\left( \bar{b}_{l+1} b_l + \bar{b}_l b_{l+1} \right)  \right] \notag \\  
 &+ \int d\tau  \left[ U \sum_{l} \left( \bar{b}_l b_l - n_0 \right)^2 + \sum_{l\neq l^{\prime}} V_{l,l^{\prime}}\bar{b}_{l} 
\bar{b}_{l^{\prime}} b_{l^{\prime}} b_{l} \right] \notag\\
&-  t^{2}_B \sum_{l} \int d\tau d\tau^{\prime} \: \bar{b}_l(\tau^{\prime})  F(\tau-\tau^{\prime})  b_l(\tau)
\label{eq:action}
\end{align}
The path integral is performed over the set of independent complex functions $b_l(\tau)$ and $\bar{b}_l(\tau)$ that obey periodic bosonic boundary 
conditions where $b_l(\tau+\beta) = b_l(\tau)$ and $\bar{b}_l(\tau+\beta) = \bar{b}_l(\tau)$, being
$\beta = T^{-1}$, i.e. the inverse absolute temperature. The above model allows to draw 
interesting parallels with a lattice system of 1D bosons moving on a periodic potential commensurate with the boson density and  coupled  to each other by weak Josephson tunneling~\cite{Ho_PhysRevLett.92.130405,Cazalilla_2006} (a dimensional crossover in  a related cold atom system has been recently experimentally observed~\cite{Guoetal2024}).
In such system, the Josephson tunneling (corresponding to dissipative coupling to the bath in the model of Eq.~\ref{eq:action}) competes with the tendency of the bosons to localize in the periodic lattice 
due to their mutual repulsion (Mott localization). This competition can be described by a set
of weak coupling renormalization-group (RG) equations that were derived and studied in Refs.~\cite{Ho_PhysRevLett.92.130405,Cazalilla_2006}. A similar set of RG equations for the 
model~\eqref{eq:action} is derived in the following section, where bosonization is employed
to obtain a low-energy effective description of the system. 
\section{Bosonization  Analysis}\label{sec:bosonization}
\subsection{Bosonized Action}

The low-energy properties of the above model can be studied using  bosonization~\cite{Cazalilla_JphysB,Cazalilla_RevModPhys.83.1405,Giamarchi_book,GogolinTsvelik1999}: The bosonic degrees of freedom described by $b_l,\bar{b}_l$ are expressed in terms of two collective
quantum fields, $\theta(x,\tau)$ and $\phi(x,\tau)$~\cite{Cazalilla_JphysB,Giamarchi_book}. The former describes local fluctuations of the phase
whilst the  spatial gradient of the latter describes the long wave-length part of the density fluctuations, i.e.
$\delta n_l/a_0 = (\bar{b}_l b_l - n_0)/a_a$ averaged over distances much larger than $a_0$, where $a_0$ is the lattice
parameter. Since the phase and the 
density are canonically conjugate variables, the fields $\theta$ and $\phi$ are dual to each other. 
In terms of these collective fields, the bosonic variables $b_l(\tau)$ can be written as 
follows~\cite{Haldane_PhysRevLett.47.1840,Giamarchi_book,Cazalilla_JphysB}:
\begin{align}
b_l(\tau) &= n^{1/2}_0 e^{i\theta(x_l,\tau)} + \cdots \\
\delta n_l(\tau) &= \frac{a_0}{\pi}\partial_x \phi(x_l,\tau) \notag\\
&\qquad +  A_1  n_0 \cos \left[ 2\phi(x_l,\tau) + 2\pi \frac{n_0 x_l}{a_0}\right] + \cdots 
\end{align}
where $x_l = l a_0$. In the above expressions the
dots correspond to other terms that have higher scaling dimensions in the RG sense~\cite{Giamarchi_book}. $A_1$ is a (dimensionless) quantity that depends on the microscopic details of the model and  therefore cannot be computed using bosonization. Introducing the above expressions into the action, Eq.~\eqref{eq:action},  the following bosonized action is 
obtained:
\begin{equation}
S[\phi,\theta] = S_B[\theta] +  S_0[\phi,\theta] + S_u[\phi] + S_D[\theta],\qquad \label{eq:act2}
\end{equation}
where
\begin{align}
S_B[\theta] &=  \int d\tau dx  \, \left[   \frac{i \delta}{2\pi}  \partial_{\tau} \theta + \frac{i}{\pi} \partial_x \phi \partial_{\tau}\theta \right], &\notag\\
S_0[\theta,\phi] &=    \int  d\tau dx \, \left[ 
\frac{vK}{2\pi} \left( \partial_{x}\theta \right)^2 + \frac{v}{2\pi K} \left( \partial_x \phi\right)^2\right],\notag\\
S_{u}[\phi] &= \frac{g_u}{\pi a_0} \int d\tau dx\:  \cos \left( 2\phi + x \delta  \right), &\notag \\
S_{D}[\theta] &=  -\frac{\alpha}{\pi a_0} \int  d\tau d\tau^{\prime}  dx\: f(\tau-\tau^{\prime}) \cos \Delta\theta(x,\tau,\tau^{\prime}), \notag\\
&\Delta\theta(x,\tau,\tau^{\prime}) = \theta(x,\tau) - \theta(x,\tau^{\prime}),\notag\\
&f(\tau, \beta \to +\infty) = \frac{1}{\tau^2} 
\end{align}
The parameter $\delta = 2\pi (n_0 - [n_0])/a_0$ (where $[n_0]$ is the integer part of $n_0$) measures the incommensurability of the lattice filling. For a model like the Bose-Hubbard model (which corresponds to Eq.~\eqref{eq:ham} with $V_{l,l^{\prime}} =0$), the  velocity $v$ and  the Luttinger  parameter $K$,  $\tilde{g}_u$  are functions of $U/t$ and cannot be obtained using the bosonization approach~\cite{Cazalilla_RevModPhys.83.1405,Giamarchi_book} but can be numerically computed using numerical methods such as Quantum Monte Carlo or density-matrix renormalization-group (DMRG), see e.g.~\cite{Cazalilla_RevModPhys.83.1405}. Assuming a weak coupling to the
bath, we have $\alpha \propto (t_B \tau_c)^2 n_0$.  This action has an associated short distance (time) cut-off $a_c\sim a_0$ ($\tau_c = a_c/v$) and therefore it describes the properties of the system in the long wave-length ($\gg a_c$) 
and long (imaginary-) time limit ($\gg \tau_c$). For integer  lattice filling $\delta = 0$, the first term in the Berry phase $S_B$ can be omitted. For general filling and in the absence of local dissipation, i.e. for $\alpha = 0$, we can integrate out the phase field $\theta$ exactly and the resulting quantum (incommensurate) sine-Gordon model describes the Mott transition in  1D  bosons systems. The latter is driven by the proliferation of phase slips~\cite{Giamarchi_book,Cazalilla_RevModPhys.83.1405} in the lattice commensurate case. For incommensurate non-dissipative systems (i.e. $\delta = 0$ and $\alpha=0$), the action describes the comensurate-incomensurate transition of the Mott insulator~\cite{Nersesyan1978,Pokrovsky1979,GogolinTsvelik1999,Giamarchi_book}.  However, in the case of dissipative systems in the continuum for which $g_u = 0$ (or for large $\delta$, see Sec.~\ref{sec:doping}), the density field $\phi$ can be 
integrated out instead. The resulting model is the dissipative Tomonaga-Luttinger liquid studied in Ref.~\cite{Cazalilla_PhysRevLett.97.076401}. 

\subsection{Renormalization Group (RG) Analysis}\label{sec:rg}
\begin{figure}[t]
    \centering
\includegraphics[width=\linewidth]{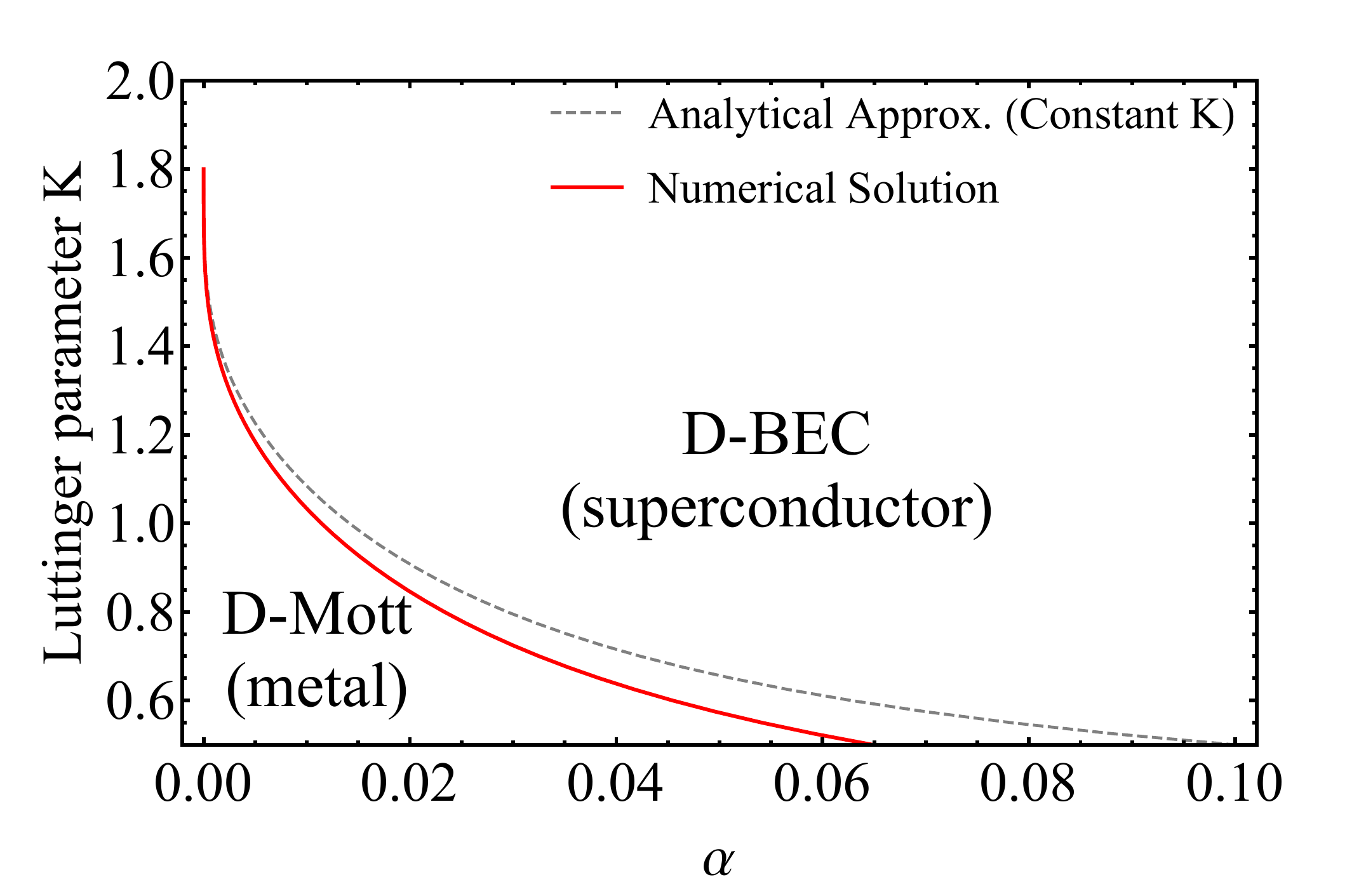}
    \caption{Phase diagram for the bosonized model~\eqref{eq:act2} for a commensurate system (i.e. $\delta = 0$) as a function of the Luttinger parameter, $K$, and the dimensionless dissipation strength, $\alpha$. A comparison is shown between the approximate solution (dashed line, obtained assuming  a constant $K$) and the full numerical solution of the renormalization-group equations \eqref{eq:rggu} to \eqref{eq:rgK}. The analytical curve is indicated by the dashed  line, while the numerical solution is represented by the thick continuous line. The line $\alpha = 0$ corresponds to a bosonic Mott insulator for $K \lesssim 2$. For the numerical and analytical solutions the following values are assumed: $g_u(0) = 2\times 10^{-3}$ and $r = 0.1$ (cf. Eq.~\ref{eq:boundary}).}
    \label{fig:fig1}
\end{figure}

In this section, we follow the approach used by Lobos \emph{et al.}~\cite{Lobos1_PhysRevB.80.214515}, who studied the bosonized action in Eqs.~\eqref{eq:act2} (which they derived for a model of a quasi-1D superconducting wire Josephson-coupled to a nearby metal electrode) using  perturbative renormalization-group (RG) methods.
The perturbative RG can be applied to the bosonized action \eqref{eq:act2} in the  limit where the couplings
$g_u$ and $g_{D}$ are weak. In addition, we also assume $\delta = 0$ (i.e. integer lattice filling). Under an RG transformation the cut-off is reduced while keeping the
partition function (up to a multiplicative constant) invariant. This process requires
the couplings in the effective action~\eqref{eq:act2} to be adjusted, and for infinitesimal
transformations the flow of the couplings is described by differential equations.
The result of this analysis yields the following set of RG equations (see Appendix~\ref{app:scaling} for details of the derivation):
\begin{align}
\frac{dg_u}{d\ell} &= (2- K) g_u,\label{eq:rggu}\\
\frac{d\alpha}{d\ell} &= \left(1- \frac{1}{2K}\right) \alpha, \label{eq:rggd}\\
\frac{dK}{d\ell} &= \alpha - \frac{g^2_u}{K},\label{eq:rgK}\\
\frac{dv}{d\ell} &=  -\frac{\alpha}{K} v. \label{eq:rgv}
\end{align}
According to the above RG equations, for  $K_{\alpha, c} = 1/2 < K  < K_{u, c}= 2$, both the Mott potential $\propto g_u$ 
and and the local phase dissipation $\propto \alpha$ are relevant perturbations to the Tomonaga-Luttinger liquid phase. In order to estimate the
boundary between the phases where either the Mott potential  or the local phase dissipation dominate, 
we have numerically solved the  RG equations to locate  the point where the renormalized
couplings satisfy the following condition:
\begin{equation}
g_u(\ell^*) = \alpha(\ell^*) = r, \label{eq:boundary}
\end{equation}
 where $r\lesssim 1$ (we have taken $r=0.1$ in the calculation
shown in Fig.~\ref{fig:fig1}). A rough estimate of the phase boundary thus obtained can be also analytically derived by neglecting the
renormalization of the Luttinger parameter $K$, and using the approximate solutions $g_u(\ell) = g_u(0) e^{(2-K)\ell}$
and $\alpha(\ell) = \alpha(0) e^{(1-K^{-1}/2)\ell}$. Imposing \eqref{eq:boundary} yields the following estimate for the
phase boundary:
\begin{equation}
\alpha(0) \sim \left[ g_u(0) \right]^{\frac{1-K^{-1}/2}{2-K}}. \label{eq:approxboundary}
\end{equation}
This estimate is compared to the numerical solution of the 
RG equations in Fig.~\ref{fig:fig1}.
Note that,  at $K \gtrsim K_{\alpha, c} = 1/2$, for which the local phase dissipation becomes marginal,  the phase boundary is especially affected by the renormalization
of  $K$. On the other hand, near $K_{u,c}=2$, for which the Mott potential becomes
marginal, the phase boundary  approaches the approximate result of Eq.~\eqref{eq:approxboundary}.
In what follows, we shall discuss the properties of these two phases and argue that the Mott 
insulating phase, for which  $ g_u(\ell)$ is the most relevant coupling, is not a conventional 1D Mott insulator (MI). However, before studying the Mott phase, we  briefly consider  the properties of the dissipative BEC phase in the next section. In this phase, the  local  dissipation is  dominant in the sense that $\alpha(\ell)$ becomes of order unity at a lower value of the RG parameter $\ell$.


\section{Dissipative BEC (D-BEC) phase}\label{sec:dbec}

In the parameter regime where the local dissipative coupling  dominates over the Mott potential, the dissipative term $S_D$ in Eq.~\eqref{eq:act2}  needs to be treated non-perturbatively. This can be done using the self-consistent
harmonic approximation as in Ref.~\cite{Cazalilla_PhysRevLett.97.076401}, which to leading order essentially implies expanding the
cosine term in $S_D[\theta]$ to second order in $\theta$, and optimizing the free energy of the resulting
quadratic action for a self-energy $\Sigma(\omega_m) \sim |\omega_m|+ \cdots$. Thus, the 
low-energy properties of the  phase are described by the following quadratic action:
\begin{align}
S_{\mathrm{SCHA}}[\theta] &= \frac{1}{2\pi \beta} \sum_{q,\omega_m} G^{-1}(q,\omega_m) |\theta(q,\omega_m)|^2,\\
G^{-1}(q,\omega_m) &=   v K \omega^2_m + K q^2/v + \Sigma(\omega_m).\label{eq:scha}
\end{align}
Notice that in SCHA action the low frequency dynamics is dissipative as it is dominated by $\Sigma(\omega_m) \sim |\omega_m|$ rather than 
$\omega^2_m$. This is because low-energy plasmons (phonons) of the original bosonic model propagate
along the chain while  carrying with them an extended cloud of quantum critical superconducting fluctuations of the bath. Put it differently, the dissipative coupling describes
the leaking by tunneling of bosons (Cooper pairs) in and out of the 1D chain. Therefore,  the Cooper pairs 
are allowed  to spend time as  pairing fluctuations in the nearby normal electrode. This `leakage' 
makes their propagation diffusive.  The diffusive dynamics has  important consequences 
for the phase correlations,
\begin{equation}
G_{\theta}(x,\tau) = \langle e^{i\theta(x,\tau)} e^{-i\theta(0,0)} \rangle,
\end{equation}
which exhibit long range order, i.e. $G_{\theta}(|x|\to +\infty, 0) = \varphi_0 = \mathrm{const.}$ and $G_{\theta}(x,|\tau|\to +\infty) =  \varphi_0$~\cite{Cazalilla_PhysRevLett.97.076401}. Thus, this phase is a Bose-Einstein condensate (BEC) with diffusive low energy excitations. In  Sec.~\ref{sec:dmott}, we shall see that the existence of this phase can be also inferred using a non-linear sigma model whose construction requires that we discuss the other phase where the Mott potential dominates. We stress that the RG treatment described above cannot provide a description of the phase transition, i.e. it cannot tell us whether it is  discontinuous or continuous and, in the former case, to which universality class it belongs. As we argue below, the TDGL  theory  introduced in~\cite{Sachdev_PhysRevLett.92.237003} and  re-derived for the present model in Sec.~\ref{sec:GL} can deal with such questions.

\section{Dissipative Mott (D-Mott) Phase}\label{sec:dmott}

In the regime where the Mott potential ($\sim g_u \cos 2\phi$ in Eq.~\ref{eq:act2}) 
is the leading relevant perturbation, based on the knowledge of the exact solution of the sine-Gordon model~\cite{GogolinTsvelik1999,Giamarchi_book} a gapped (Mott insulator, MI)  phase is expected to be stabilized. The MI has a spectral gap and therefore the  phase correlations $G_{\theta}(x,\tau) = 
\langle e^{i\theta(x,\tau)}  e^{-i\theta(0,0)}\rangle$ decay exponentially both in space $x$ and imaginary time $\tau$.  However, in the
presence of local phase dissipation, it is shown below that this picture is not accurate.
Instead,  the phase that is stabilized in the presence of both Mott potential and local phase dissipation has some distinct features and we shall refer to it below as "dissipative Mott"  (D-Mott) phase or, simply metal.

\begin{figure}[t]
    \centering
\includegraphics[width=\linewidth]{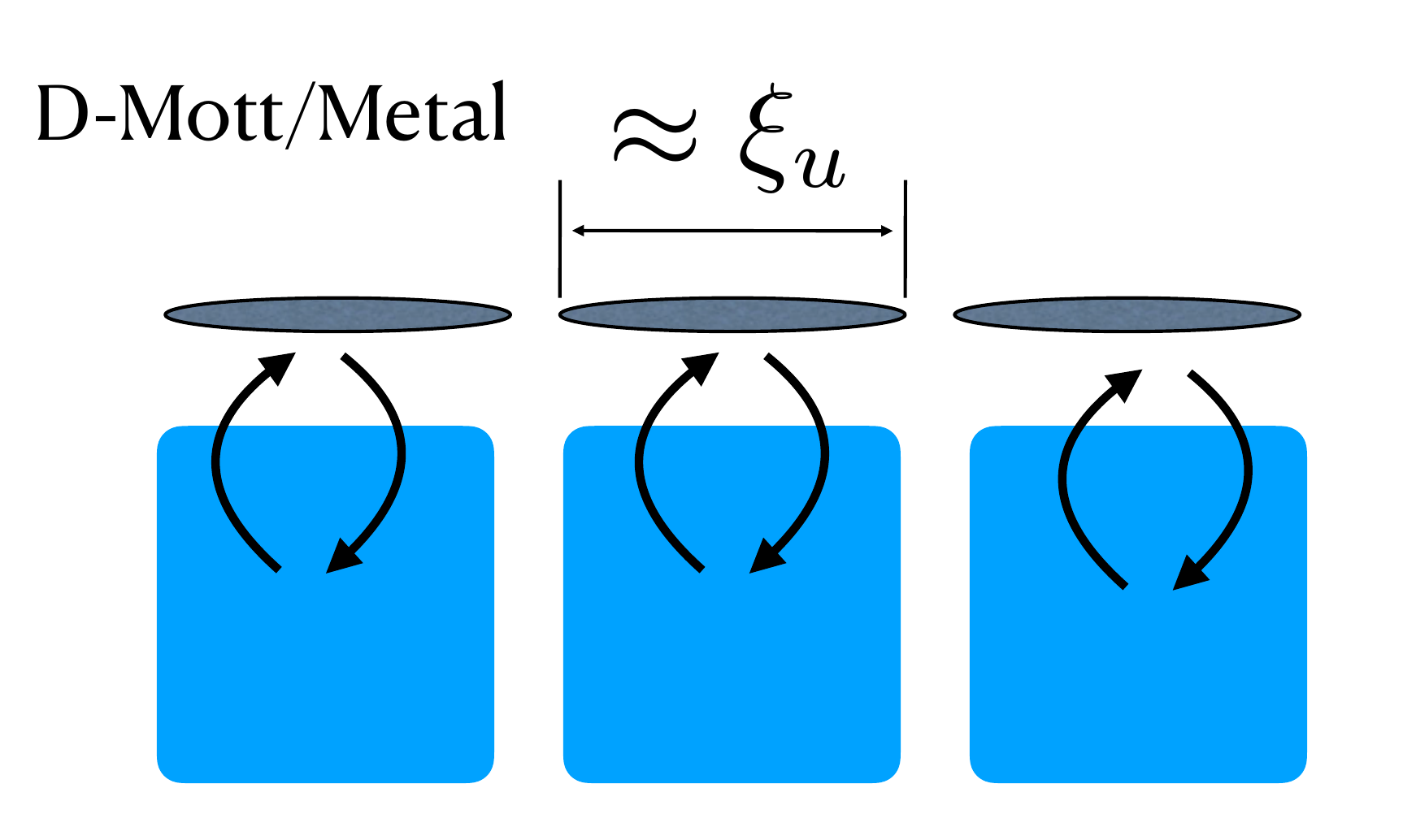}
    \caption{Sketch of the D-Mott/Metal phase after coarse-graining the system: The 1D boson system + bath breaks into an array of weakly Josephson-coupled bosonic puddles that are coupled to individual Ohmic  baths. The latter collectively represent a nearby diffusive metal electrode. The bosons (Cooper pairs of a 1D superconductor) can leak in and out of the puddles as indicated by the arrows. The characteristic size of the puddles $\approx \xi_u$ is (roughly) determined by the length scale beyond which phase slips caused by the underlying lattice suppress coherent boson tunneling (see discussion in Sec.~\ref{sec:dmott}).}
    \label{fig:puddles}
\end{figure}

 The perturbative RG calculations discussed in Sec.~\ref{app:scaling} show that, 
 when the bare couplings $g_u(0), \alpha(0)$ are such that $g_u(\ell)$ becomes of order one at a smaller value of $\ell = \ell_u$, i.e.  at  a length scale $\xi_u = a_0 e^{\ell_u}$ such that $g_u(\ell_u)\sim 1$, the phase slips strongly suppress boson hopping  beyond distances $\gtrsim \xi_u$. However, although
$g_u(\ell_u) > \alpha(\ell_u)$ in this parameter regime, the local phase
dissipation is still a  \emph{subleading} relevant perturbation and must be 
taken care of. This can be achieved as follows: Roughly speaking, $\xi_u$ can be
thought of as the length scale beyond which quantum coherent tunneling is strongly suppressed
 due to  phase slips. Notice that $\xi_u$ defined in this way (e.g. from solution to RG equations of the previouss section) does not diverge even near the  phase boundary estimated in Sec.~\ref{sec:rg} where $\ell_u = \ell^*$ and $g_u(\ell^*) = \alpha(\ell^*) \lesssim 1$. The lack of a diverging length scale may  be interpreted as a failure of the perturbative RG approach  to describe the nature of the transition, which, as we shall see below, is a continuous one. However, in the present case we shall instead  adopt the 
viewpoint that $\xi_u$, as estimated from the RG equations, corresponds to the coherence length for boson hopping. Beyond the scale of $\xi_u$, the D-Mott phase be  regarded as an incoherent array of one-dimensional boson ``puddles',  each  coupled to a local ohmic bath (see Fig.~\ref{fig:puddles}).
Mathematically, it is possible to arrive at such description if we coarse-grain the system and rewrite the action~\eqref{eq:action}  as an 1D array of Josephson
junctions with onsite ohmic  dissipation.  Thus, assuming that the number of particles per puddle  (site)  $N_0$ is large, we  express
the  boson fields $b_l(\tau), \bar{b}_l(\tau)$ as follows:
\begin{align}
b_l(\tau) = \left(N_0 +\delta n_l(\tau)\right)^{1/2}\:  e^{+i\theta_l(\tau)},\\
\bar{b}_l(\tau) = \left(N_0 +\delta n_l(\tau)\right)^{1/2}\: e^{-i\theta_l(\tau)}.\label{eq:phasedens}
\end{align}
In this limit, the particle number fluctuations of the puddle $\delta n_l$ are suppressed by its charging energy $E_C \approx  U$, i.e. $|\delta n_l(\tau)| \ll N_0$,  and therefore they  can be integrated out leading to following model:
\begin{align}
S\left[ \theta_l \right] &=  S_B[\theta_l] +  S_C[\theta] +    S_{D}[\theta_l] + S_{J}[\theta_l], \label{eq:act3}\\
S_{B}[\theta_l] &= i  N_0 \sum_{l} \int d\tau \, \partial_{\tau}\theta_l,\notag\\
S_C[\theta] &= \sum_{l} \int d\tau \frac{\left( \partial_{\tau}\theta_l\right)^2 }{2E_C},  \notag\\
S_{D}[\theta_l] &=   - \frac{\alpha_p}{2} \sum_l \int d\tau d\tau^{\prime} f(\tau-\tau^{\prime}) \cos\left[\theta_l(\tau) - \theta_{l}(\tau^{\prime})\right],\notag\\
S_J[\theta_l] &=  - J \sum_{l}  \int d\tau \cos\left[\theta_l(\tau) -\theta_{l+1}(\tau)\right],\notag\\
&f(\tau,\beta\to +\infty) = \frac{1}{\tau^2},
\label{eq:act33}
\end{align}
where  the puddle dissipation strength $\alpha_p  \approx (t_B \tau_c)^2 N_0 \propto \alpha$. 
This (coarse-grained) description applies to  a 1D array of boson puddles of size $\approx \xi_u$ (cf. Fig.~\ref{fig:puddles}).  The latter  contain many bosons, unlike the sites of the original lattice boson model, Eq.~\eqref{eq:action}, for which 
$n_0$ may well be of order unity (i.e. $N_0 \gg n_0 \sim 1$).  Therefore, the assumption of $N_0\gg 1$ is well justified. For a commensurate lattice of puddles, the filling $N_0$ should be treated as an integer and therefore, as in the bosonized case Eq.~\eqref{eq:act2}, the Berry phase $S_B[\theta_l]$ can be omitted. Nevertheless, it is important to stress that the Berry phases in   Eq.~\eqref{eq:act2} and \eqref{eq:act33} keep track of the (in)commensurability and in general they cannot be ignored. Thus, for instance, $N_0$ will be (half-) integer when $n_0$ is (half-) integer. We stress this point because it will be important for the discussion of the doping of the D-Mott phase in Sec.~\ref{sec:doping}.  

In Eq.~\eqref{eq:act33} $J \approx t N_0$ is the Josephson energy. In the derivation of (\ref{eq:act3},\ref{eq:act33}), we  have assumed that $V_{l,l^{\prime}} = 0$. Accounting for $V_{l,l^{\prime}}\neq 0$ leads to an additional term of the form $C_{l,l^{\prime}}\partial_{\tau} \theta_{l}\partial_{\tau}\theta_{l^{\prime}}$, which we assume to be negligible due to the  screening of long-range interactions provided by the diffusive metal electrode. The picture of a 1D array of dissipative puddles coincides with the description of an array of superconducting islands in the vicinity of a dirty metal studied in Refs.~\cite{FEIGELMAN1998107,Larkin_PhysRevLett.86.1869}. Thus,  
after coarse-graining the present dissipative boson system, the distinction between a single-channel  (i.e. purely 1D) wire and granular system blurs.

 In order to further understand the properties of the  D-Mott phase, we start by neglecting the Josephson coupling, i.e. setting $J = 0$ in Eq.~\eqref{eq:act3}. The resulting model is an array of decoupled dissipative quantum rotors described by the terms $S_C[\theta_l] + S_D[\theta_l]$ (cf. 2nd and 3rd lines of Eq.~\ref{eq:act33}). Indeed, the properties of the  dissipative quantum rotor model  have been  studied extensively (see e.g. ~\cite{Renn2,Grabert_PhysRevLett.81.2324,Hofstetter_PhysRevLett.78.3737,Spohn1999,Lukyanov_2006,LeDoussal_PhysRevB.82.155127} and references therein) and we can use the following exact result in order to treat the Josephson coupling perturbatively: For a  single dissipative quantum rotor,  the local phase correlations at zero temperature asymptotically behave as follows~\cite{Spohn1999}:
\begin{equation}
G_{\theta}(\tau) = \langle e^{i\theta_l(\tau)} e^{-i\theta_l(0)}\rangle \simeq \mathcal{A} \left(\frac{\tau_c}{\tau}\right)^2, \label{eq:spohn}
\end{equation}
where $\mathcal{A} = \mathcal{A}(\alpha_p)$ is a non-universal prefactor and $\tau\gg \tau_c$. This result can be also
obtained by generalizing the  dissipative rotor model from $O(2)$ to $O(N)$ symmetry and taking the large-$N$ limit~\cite{Renn1,Renn2} (see Appendix~\ref{app:largeN} for a review of these results), and from perturbation theory~\cite{LeDoussal_PhysRevB.82.155127} for small $\alpha_p$. Using Eq.~\eqref{eq:spohn}, we obtain, to leading order in $J$, the following results: (i) The correction to ground state energy is finite, and  (ii) the asymptotic equal-time phase correlations  decay  exponentially with distance.

 In order to show (i),  consider the 
 leading-order correction, $\Delta E_0$, in powers of $J$ 
 to the ground state energy $E_0$ for Eq.~\eqref{eq:act3}(see also Appendix~\ref{app:pertchi}):
\begin{align}
\Delta E_0 &= -\lim_{\beta\to +\infty} \left(\frac{J^2}{2!\beta}\right) \sum_{l,l^{\prime}} \int d\tau d\tau^{\prime} \langle \cos\left[\theta_l(\tau) - \theta_{l+1}(\tau)\right] \notag\\ 
&\qquad \times\cos\left[\theta_{l^{\prime}}(\tau^{\prime}) - \theta_{l^{\prime}+1}(\tau^{\prime})\right] \rangle_0  + O(J^4).
\end{align}
Here $\langle \ldots \rangle_{0}$ denotes average with respect to the action in Eq.~\eqref{eq:act3} with $J=0$.
Note that only even orders in $J$ appear in the above expression due to the ``local gauge invariance'' of
Eq.~\eqref{eq:act3} at $J = 0$,  which implies that any correlations must be 
invariant under $\theta_l(\tau) \to \theta_l(\tau) + \theta^0_l$, where $\theta^0_l$ is a constant that is different at each site.  Using \eqref{eq:spohn} 
\begin{align}
 &\langle e^{\pm i\left[\theta_l(\tau) - \theta_{l+1}(\tau)\right]} e^{\mp i\left[\theta_{l^{\prime}}(\tau^{\prime}) - \theta_{l^{\prime}+1}(\tau^{\prime})\right]} \rangle_{0} \notag\\
 &\qquad= \left[ G^0_{\theta}(\tau-\tau^{\prime})\right]^2\delta_{l,\l^{\prime}} \simeq  \mathcal{A}^2 \left(\frac{\tau_c}{\tau-\tau^{\prime}} \right)^2 \delta_{l,l^{\prime}}
\end{align}
for $|\tau-\tau^{\prime}| \gg \tau_c$. Upon integrating over $\tau$, the above expression yields a finite $O(J^2)$ correction to ground state energy. Higher-order corrections in $J$ involve higher-order correlation functions of  the phase $e^{\pm i\theta_l(\tau)}$ for which, to the best of our knowledge,
no exact expressions  are known.  
Nevertheless, noticing that  the large-$N$ approach  reproduces~\eqref{eq:spohn} and  the theory 
becomes gaussian in this limit (see Appendix~\ref{app:largeN}),  higher order correlation functions can be  approximated by products of two-point correlation functions. The latter decay sufficiently fast as power-laws of $\tau$ and therefore yield convergent corrections to the ground state energy. 

Next we take up on the derivation of result (ii).  It follows from the leading perturbative correction for the equal-time phase correlator:
\begin{multline}
\langle e^{i\theta_{l}(0)} e^{-i\theta_{0}(0)}\rangle =   \left(\frac{J}{2} \right)^{l} \int d\tau_{l} \cdots  d\tau_1\, \\
\qquad \times  \langle e^{i\theta_l(0)} e^{-i \theta_l(\tau_{l})} \rangle_{0} \times \langle e^{i\theta_{l-1}(\tau_{l})} e^{-i \theta_{l-1}(\tau_{l-1})} \rangle_{0}  \\
\times\cdots\times \langle e^{i\theta_{0}(\tau_1)} e^{-i \theta_0(0)} \rangle_{0} 
 + O(J^{l+2})\\
=  J^{l} \int \frac{d\omega_m}{(2\pi)} \left[ G_{\theta}(\omega_m)\right]^{l+1} + O(J^{l+2}),
\end{multline}
The last integral can be estimated by using $G^{0}_{\theta}(\omega_m) \simeq \mathcal{A} \tau^2_c |\omega_m|$, which is the Fourier transform of  Eq.~\eqref{eq:spohn}. Hennce, 
\begin{align}
G^{J\neq 0}_{\theta}(x_l,0) &= \langle e^{i\theta_{l}(0)} e^{-i\theta_{0}(0)}\rangle \notag\\
 &\approx \mathcal{A}\tau_c ( \mathcal {A} J \tau_c)^{l} \: \int^{\tau^{-1}_c}_0 \frac{d\omega_c}{\pi} \, |\omega_m\tau_c|^{l+1} \\
&\sim \frac{e^{-|x_l|/\xi_c},}{|x_l|} 
\end{align}
where $a_p \xi^{-1}_c \approx  - \log \left(J \tau_c \mathcal{A} \right)$ and $x_l =a_p l$.  
Thus, to leading order in $J\tau_c$, equal-time phase correlations exhibit a strong exponential decay  with distance. On the other hand, the local phase correlations are given by Eq.~\eqref{eq:spohn}:
\begin{align}
G^{J\neq 0}_{\theta}(x_l = 0,\tau) &=  \langle e^{i\theta_{0}(\tau)} e^{-i\theta_{0}(0)}\rangle\notag\\ 
&= G_{\theta^{0}}(\tau)  \sim \left( \frac{\tau_c}{\tau} \right)^2 + O(J^2).
\end{align}
This behavior is not expected for the conventional MI  described  by the sine-Gordon model (in the continuum limit) or the  $U \gg t$ limit by the Bose-Hubbard model~\cite{Giamarchi_book}. This is because the local dissipation substantially
modifies the phase correlation properties of the MI in imaginary time changing an exponential decay into a power-law. Indeed, 
the spatial localization of the bosons caused by interactions in 1D is compatible with some
degree of local phase ordering in imaginary time caused by the  delocalization of bosons  in the dissipative bath. In this sense, this phase can be regarded as an array of dissipative boson puddles (cf. Fig.~\ref{fig:puddles}).  

The above perturbative treatment breaks down at a finite strength of the Josephson coupling $J$. Indeed, for the MI in the absence of dissipation,  it  yields an exponential decay of the phase correlations (in this case the local phase correlations  decay exponentially in imaginary time since the decoupled rotors are described by the $O(2)$ quantum rotor model in $0+1$ dimensions, see Appendix~\ref{app:freecorr}).  The regularity of the perturbative series is a consequence of adiabatic continuity:  The  D-Mott phase is adiabatically connected to an array of  independent \emph{dissipative} $O(2)$ rotors in $0+1$ dimensions. However, perturbation theory cannot anticipate the existence of a phase transition
to the ordered D-BEC phase. The latter  arises from singularities in the partition function that
are not accessible through perturbation theory.

\section{Continuum limit of puddle model}\label{sec:crotor}

The picture of the D-Mott/metal phase provided above appears to be  accurate 
for large ratio $U/t$ in the original boson model. However,  if the hopping amplitude, $t$, of the  bosons is increased, the Luttinger parameter in Eq.~\eqref{eq:act2} also increases $K$~\cite{Cazalilla_RevModPhys.83.1405}. Concurrently, the dimensionless Mott potential $g_u(0)  \sim U/t$,  decreases. Using the approximate phase boundary obtained from the weak-coupling RG equations (cf. Sec.~\ref{sec:rg}), the 
estimated  value for the ``critical'' dissipation strength $\alpha_c$ shifts to lower values of $\alpha$, i.e. $\alpha_c = \alpha(0)  \sim \left[ g_u(0) \right]^{\frac{1-K^{-1}/2}{2-K}}$, see also Fig.~\ref{fig:fig1}. For the coarse-grained system of puddles, 
this translates into a larger ratio $J/E_C$ and an overall weaker dissipation strength  for the  transition to the D-BEC phase.  In this situation,  due to a large $J$ the phase is expected to fluctuate less from puddle to puddle and, therefore,  we may take the \emph{na\"ive} continuum limit of the dissipative rotor (puddle) model, Eqs.~\eqref{eq:act3} and \eqref{eq:act33}. In the 
absence of dissipation (i.e. for $\alpha_p = 0$,) this procedure  ignores the existence of  phase slips and  the superconductor-(Mott) insulator transition~\cite{GogolinTsvelik1999}. However, as we show in this section,  the existence of a local phase dissipation leads to a completely different picture, which agrees with  the results  of both the SCHA described in Sec.~\ref{sec:dbec} for the D-BEC/superconductor,  and the perturbative results of Sec.~\ref{sec:dmott} for the D-Mott/metal. As a bonus, an analysis of the resulting continuum model in the large-$N$ limit provides a reasonable estimate of the critical exponents at the quantum phase transition.

 In order to take the \emph{na\"ive} continuum limit of Eqs.~\eqref{eq:act3} and \eqref{eq:act33},   it is convenient to introduce the following two-component unit vector:   $\vec{n}_l(\tau) = \left( \cos \theta_l(\tau), \sin \theta_l(\tau) \right)$, which allows us to rewrite the action of the dissipative Josephson-junction array (at commensurate filling) as follows:
 \begin{align}
S[\vec{n}_l] &= S_D[\vec{n}_l] + S_{J}[\vec{n}_l]\notag\\
S_D[\vec{n}_l] &= \frac{1}{2\beta} \sum_{l} \sum_{\omega_m} \left[ \alpha_p |\omega_m|   + \frac{\omega^2_m}{E_C} \right] |\vec{n}_l(\omega_m)|^2, \notag\\
S_J[\vec{n}_l] &= \frac{J}{2} \sum_{l} \int d\tau \left[ \vec{n}_l(\tau) - \vec{n}_{l+1}(\tau) \right]^2. \label{eq:coupledrotors}
 \end{align}
 Next we define an interpolating vector field, $\vec{n}(x,\tau)$,
 such that $\vec{n}(x = x_l, \tau) = \vec{n}_l(\tau)$, where $x_l = l a_p$  ($a_p \approx \xi_u$ is the puddle array lattice parameter, with $\xi_u$ 
   the puddle-size or correlation length introduced  in Sec.~\ref{sec:dmott}). 
  After performing a gradient expansion, the following  NL$\sigma$M in $1+1$ dimensions is obtained:
 \begin{align}
 S[\vec{n}] &= \frac{1}{2\beta L} \sum_{q,\omega_m} G(q,\omega_m) |\vec{n}(q,\omega_m)|^2,\\
 G^{-1}(q,\omega_m) &= \eta |\omega_m| +  \frac{\omega^2_m}{\gamma}   + \kappa q^2, \label{eq:cNLsM}
 \end{align}
 where $\vec{n}(q,\omega) = \int dx d\tau e^{i q x - i\omega_m \tau} \vec{n}(x,\tau)$.
 Here $\eta = \alpha_p/a_p$, $\kappa   \sim J a_p$, and $\gamma \sim a_p E_C$. 
 To above action we must  add  a Lagrange multiplier termp $\propto \lambda(x,\tau)$
 in order to en the constraint $\left[\vec{n}(x,\tau)\right]^2 = 1$ at every point $(x,\tau)$.
 Notice that, although this model is  formally identical to the NL$\sigma$M introduced in Ref.~\cite{Cazalilla_PhysRevLett.97.076401},  in this case the continuum
 limit is taken starting from a proper lattice model and not from the bosonized
 action (see Sec.~\ref{sec:concl} for further discussion of this topic). 
 
 We can analytically study the above NL$\sigma$M by generalizing the  symmetry 
 from O$(2)$ to O$(N)$ following Refs.~\cite{Renn1,Renn2,Cazalilla_PhysRevLett.97.076401},
 In the large-$N$ limit, the model becomes a
 Gaussian field theory. The reader is referred to  Appendix~\ref{app:largeN} 
 for the details of the calculations. Below,  a summary of the most important results concerning the phase properties is
 provided. The large-$N$ approach predicts a phase transition as function of the strength of the dissipation $\eta\propto \alpha_p (\alpha)$. In the weak dissipation regime, the phase is disordered at zero temperature. 
 This is manifested in the correlations of the vector field $\vec{n}(x,\tau)$ at $T = 0$ asymptotically taking the following limiting forms:
 \begin{align}
G_{\theta}(x,0) &=  \langle \vec{n}(x,0)\cdot \vec{n}(0,0)\rangle \sim e^{-|x|/\xi_c},\\
 G_{\theta}(0,\tau) &= \langle \vec{n}(0,\tau) \cdot \vec{n}(0,0)\rangle \sim \frac{1}{\tau^2}.
 \end{align}
 where the (renormalized) correlation length diverges as $\xi_c/a_p  = (\eta -\eta^*)^{-1}$ close to the transition. On the other hand, 
 in the ordered phase at large dissipation, the asymptotic (phase) correlations at zero temperature 
 take the form:
  \begin{align}
  \lim_{|x|\to +\infty}\langle \vec{n}(x,0)\cdot \vec{n}(0,0)\rangle  = C^2_0 + \ldots,\\
\lim_{|\tau|\to+\infty}\langle \vec{n}(0,\tau) \cdot \vec{n}(0,0)\rangle = C^2_0 + \ldots
 \end{align}
 where $C_0$ is a (non-universal constant) and the ellipsis stands for  corrections that decay as power-laws (see Appendix~\ref{app:largeN}). 
 Thus, as mentioned above,  the NL$\sigma$M is able to reproduce the results of both the SCHA and the perturbative approach in the limit of strong and weak dissipation, respectively.

 \section{Dissipative TDGL theory}\label{sec:GL}

 In this section, starting from the puddle picture introduced above, we show that the above $1+1$ NL$\sigma$M is indeed equivalent to the dissipative time-dependent Ginsbug-Landau  (TDGL) theory used by  Refs.~\cite{Sachdev_PhysRevLett.92.237003,Werner_JPSP} to describe the superconductor-metal transition in quasi-1D superconductor wires.
 Below we provide a microscopic derivation of the dissipative TDGL  in the present context, which essentially amounts to replacing the quantum rotor field $\vec{n}(x,\tau)$ with its hard  unit-length constraint by a ``soft vector'' field $\vec{\Phi}(x,\tau)$ subject to a Mexican-hat potential.  The  derivation is  most conveniently carried out starting from the coupled rotor  model, Eq.~\eqref{eq:coupledrotors}, and carrying out a Hubbard-Stratonovich (HS) transformation with the  HS vector  $\vec{\Phi}_l(\tau)$ coupled to $\vec{n}_l(\tau) = \left( \cos \theta_l(\tau), \sin\theta_l(\tau) \right)$. The transformation  decouples the (Josephson) hopping term in Eq.~\eqref{eq:coupled} and leaves us with a set of  decoupled dissipative rotors with the $\vec{\Phi}_l(\tau)$ as external source fields. Integrating out the rotor variables $\vec{n}_l(\tau)$ leads to the following effective action for  $\vec{\Phi}_l(\tau)$:
\begin{equation}
e^{-S_D[\vec{\Phi}_l]} =  \prod_{l} \int    \left[ d\vec{n_l} \right]  e^{-S_D[\vec{n}_l] - \int d\tau \vec{\Phi}_l \cdot \vec{n}_l} 
\end{equation}
In practice, the integration over $\vec{n}_l$ is carried out using the cumulant expansion~\cite{Lobos_PhysRevB.86.035455,Refael_PhysRevLett.101.106402}. 
To leading order in powers of $\vec{\Phi}_l(\tau)$ we obtain:
\begin{align}
S_{D}[\Phi_l] &= \frac{1}{2}\sum_{l} \int d\tau_1 \tau_2\: G_{\theta}(\tau_1-\tau_2) \vec{\Phi}_{l}(\tau) \cdot \vec{\Phi}_{l}(\tau_2) \notag\\
 &+ \frac{u}{4} \sum_{l} \int d\tau \left[ \vec{\Phi}_{l}(\tau) \cdot \vec{\Phi}_{l}(\tau)\right]^2 + \dots
 \end{align}
Here 
\begin{equation}
u \propto \int \prod_{i=1}^4 d\tau_{i}\, \langle \mathcal{T} \left[ e^{i\theta_l(\tau_1)}  e^{i\theta(\tau_2)} e^{-i\theta(\tau_3)} e^{-i\theta(\tau_4)}\right]\rangle_c \, ,
\end{equation}
where $\langle \ldots \rangle_c$ stands for the cumulant (i.e. ``connected'')  average of $e^{\pm i\theta(\tau)}$.
Taking into account the asymptotic behavior of $G_{\theta}(\tau_1-\tau_2)$ shown in Eq.~\eqref{eq:spohn}, the second order term can be written
as
\begin{equation} 
\frac{1}{2\beta} \sum_{l,\omega_m}  \left[ u_2 + z_1 |\omega_m| \right] |\vec{\Phi}(\omega_m)|^2
\end{equation}
for small $\omega_m$. The Hubbard-Stratonovich variables $\vec{\Phi}_l(\tau)$ are coupled by means of
\begin{equation}
S_J[\vec{\Phi}_l] =   \sum_{l,l^{\prime}} \int d\tau\,   (J^{-1})_{l,l^{\prime}} \vec{\Phi}_{l}(\tau) \cdot \vec{\Phi}_{l^{\prime}}(\tau),  
\end{equation}
where $J^{-1}$ is the inverse of the  matrix $J_{l,l^{\prime}} = \tfrac{1}{2}  \left( \delta_{l,l^{\prime}+1} + \delta_{l,l^{\prime}-1} \right)$. 
This term is best dealt by expanding $\vec{\Phi}_l$  in a Fourier series:
\begin{equation}
\vec{\Phi}_{l}(\tau) = \frac{1}{\beta L} \sum_{q,\omega_m} e^{i q x_l -i\omega_m \tau}  \
\vec{\Phi}(q,\omega_m).
\end{equation}
Hence, in the small $q$ limit $S_J\left[\vec{\Phi}\right]   \propto \tfrac{1}{2}\left( J \beta L\right)^{-1} \sum_{q,\omega_m} \left[ 1 + (qa_p)^2/2 \right] \: |\vec{\Phi}(q,\omega_m)|^2$. Rescaling/relabeling $q$ as $Q$  and $\vec{\Phi}$ the following dissipative field theory is obtained in the thermodynamic limit~\cite{Pankov_PhysRevB.69.054426}:
\begin{align}
S[\vec{\Phi}] &= \frac{1}{2\beta} \sum_{\omega_m} \int \frac{dQ}{2\pi} \left[ r +  |\omega_m| +  Q^2 \right] |\vec{\Phi}(Q,\omega_m)|^2 \notag\\
& \qquad+ \frac{u}{4!} \int d\tau dx  \left[ \vec{\Phi}^2(x,\tau)  \right]^2. \label{eq:tdgl}
\end{align}
This dissipative TDGL theory describes the BEC phase for $r  \sim u_2 - J^{-1}/2 < 0$, for which the uniform part of the action is minimized by any constant vector $\vec{\Phi}$ such $|\vec{\Phi}| = \Phi_0 = \sqrt{- 6 r/u} \neq 0$. By studying  the (Gaussian) fluctuations  about the ordered state where $\vec{\Phi}(x,\tau) = \Phi_0 \left(\cos \varphi(x,\tau),  - \sin \varphi(x,\tau) \right)$ for $\varphi(x,\tau)\ll 1$,  the effective action   of  the ``Goldstone'' field $\varphi(x,\tau)$ takes the same form as the one obtained using the SCHA (cf. Eq.~\eqref{eq:scha}  in Sec.~\ref{sec:dbec}).  In a more general context, the theory in Eq.~\eqref{eq:tdgl} is the $O(2)$ Hertz-Millis-Moriya theory in $1+1$ dimensions, which has been extensively studied in the context of quantum critical phenomena, see e.g. \cite{Pankov_PhysRevB.69.054426} and references therein.

 The above NL$\sigma$M theory can also  the describe the quantum criticality of the transition to the D-Mott phase, which belongs to the  Wilson-Fisher universality class~\cite{Pankov_PhysRevB.69.054426,Werner_JPSP} and  is therefore  different from  BKT universality class transition of  the transition from  the gapless TLL (superconductor) to the  1D MI~\cite{Giamarchi_book,Cazalilla_RevModPhys.83.1405}. As mentioned in the introduction, the transition was studied numerically in Ref.~\cite{Werner_JPSP} and the critical exponents obtained from the NL$\sigma$M numerically not far from the numerical values obtained from the Monte Carlo calculations and the $\epsilon$-expansion results reported in Ref.~\cite{Pankov_PhysRevB.69.054426} (see Appendix~\ref{app:largeN} for additional details). 

\section{Doping the D-Mott phase}\label{sec:doping}

 The MI phase of e.g. the Bose-Hubbard model has a spectral gap which makes it incompressible. In one dimension, the chemical potential $\mu$ must overcome this gap to drive the system into a Tomonaga-Luttinger liquid phase through a commensurate-incommensurate transition~\cite{Nersesyan1978,Pokrovsky1979,GogolinTsvelik1999,Cazalilla_RevModPhys.83.1405}. Thus, for  $\mu > \mu_c$,  defects in the MI taking the form of `holes' (absence of a boson) or 'particles' (excess of a boson) are introduced, eventually destroying the Mott insulating state and its gap.
At a low density of defects, the holes or particles behave as a one-dimensional hard-core Bose gas. As the density of defects is further increased, the residual interactions between the defects result in  algebraically decaying correlations with interaction-dependent exponents~\cite{Haldane_PhysRevLett.47.1840,Giamarchi_book,Cazalilla_RevModPhys.83.1405}.

 However, as described in Sec.~\ref{sec:dmott}, the D-Mott phase that is stabilized by the quantum dissipation has different properties from a 1D MI phase.
 By relying on the picture of the D-Mott phase as a system of weakly coupled boson puddles,  we show in what follows that the D-Mott phase has a finite compressibility. This can be seen  by using perturbation theory in both the dissipation strength and the Josephson coupling.  Before getting into the details of the calculation, it is important to notice that, for non-integer puddle filling $N_0$, the integer part of $N_0$ (i.e. $[N_0]$)  has no physical consequences (see the Berry phase term  in Eq.~\eqref{eq:act33} and  also Appendix~\ref{app:pertchi}). Thus, in following  discussion, when writing $N_0$ we actually mean $N_0 - [N_0]$ and from this point on we shall  assume that $|N_0| \leq \tfrac{1}{2}$.  
  
We have computed the zero-temperature compressibility in the weak $J$ and $\alpha_p$ limits by means of a perturbative approach similar to the one used in Sec.~\ref{sec:dmott} for the ground state energy.  The result takes the form (see  Appendix~\ref{app:pertchi} for full details):
\begin{align}
\chi_0(N_0) &= -\frac{\partial^2 (E_0/N_P)}{\partial \mu^2} =  \frac{ \alpha_p}{E_C}  \left[  \mathcal{C}_1(N_0)  + \frac{J^2}{2E^2_C} \mathcal{C}_2(N_0) \right],\notag\\
\mathcal{C}_1(N_0) &=  \frac{2}{1-4 N^2_0},\notag\\ 
\mathcal{C}_3(N_0) &= 32 \left[ \frac{15 + 152N_0^2 - 80N_0^4}{(1 - 4N_0^2)^3 (25 - 4N_0^2)} \right],\label{eq:compress}
\end{align}
where $N_P \sim L$ is the number of puddles. 
For $|N_0| < \tfrac{1}{2}$ the coefficients  $\mathcal{C}_{1,2}(N_0)$ are positive definite and monotonically increasing functions of $|N_0|$, i.e. they fulfill $\mathcal{C}_{1,2}(|N_0| > 0) >  \mathcal{C}_{1,2}(N_0 = 0)> 0$, with $\mathcal{C}_1(0) = 2$ and $\mathcal{C}_2(0) = 96/5$.  The divergence at $|N_0| = 
\tfrac{1}{2}$ is caused by the breakdown of perturbation theory due to the degeneracy of the ground state at half-integer filling~\cite{Matveev1991,Grabert_PhysRevLett.81.2324} (see  below).  

In the decoupled-puddle regime (i.e. for $J = 0$), the compressibility of a single dissipative rotor  has been previously computed using both 
analytical and numerical  (Monte Carlo) techniques by a number of authors~\cite{Hofstetter_PhysRevLett.78.3737,Grabert_PhysRevLett.81.2324,Lukyanov_2006}. Our $O(\alpha_p)$ result in this limit agrees with the results reported earlier for $N_0$ in Ref.~\cite{Hofstetter_PhysRevLett.78.3737} and for $N_0\neq 0$ in Ref.~\cite{Grabert_PhysRevLett.81.2324}. In addition, the O($\alpha_p J^2/E_C)$ correction obtained here further increases the compressibility, which is  sensible since the Josephson coupling of the puddles provides extra channels for the bosons to delocalize. Being perturbative in the dissipation strength, $\alpha_p$, the result in Eq.~\eqref{eq:compress} is relevant to the stability of the doped D-Mott phase, which appears as the favored ground state in the weak dissipation regime of the model.

Eq.~\eqref{eq:compress} implies that, at least  for small changes in the lattice filling around integer filling (corresponding to $N_0 = 0$ here), a weakly coupled system of puddles can absorb the excess of particles or holes introduced by doping. Indeed, if we use the bosonized action, Eq.~\eqref{eq:act2},  the characteristic length scale of the incommensurability, i.e. $\sim \delta^{-1}$~\cite{Giamarchi_book}, is larger than the puddle size $\approx \xi_u$ at  small  $\delta$. Furthermore,  the delocalization of the bosons between the 1D lattice and the  dissipative baths causes    particle-number fluctuations  of the puddles that result in the finite compressibility of Eq.~\eqref{eq:compress}. In a sense, the finite compressibility is not surprising because, in the presence of dissipation, the  total boson number of the chain is no longer a conserved quantity and can undergo fluctuations. The puddle array can accommodate the excess of particles (or holes)  introduced into the system as long as $\xi_u \lesssim \delta^{-1}$.  This situation is very different from a 1D Bose MI for which $N$ is strictly conserved and  bosons  are localized  by interactions around their lattice sites, which leads to   strongly suppressed  number fluctuations at low-energies and an abrupt response to the addition or removal particles. The local quantum  dissipation  softens this response ensuring the stability of the D-Mott phase  at small doping. Therefore, based on our previous considerations, in the doped case the quantum criticality of the transition  from the D-Mott to the D-BEC phases  driven by increasing $\alpha\sim \alpha_p$ or the boson tunneling amplitude $t\sim J$  is expected to belong to the Wilson-Fisher universality class as in the commensurate case. 
\begin{figure}[t]
    \centering
\includegraphics[width=\linewidth]{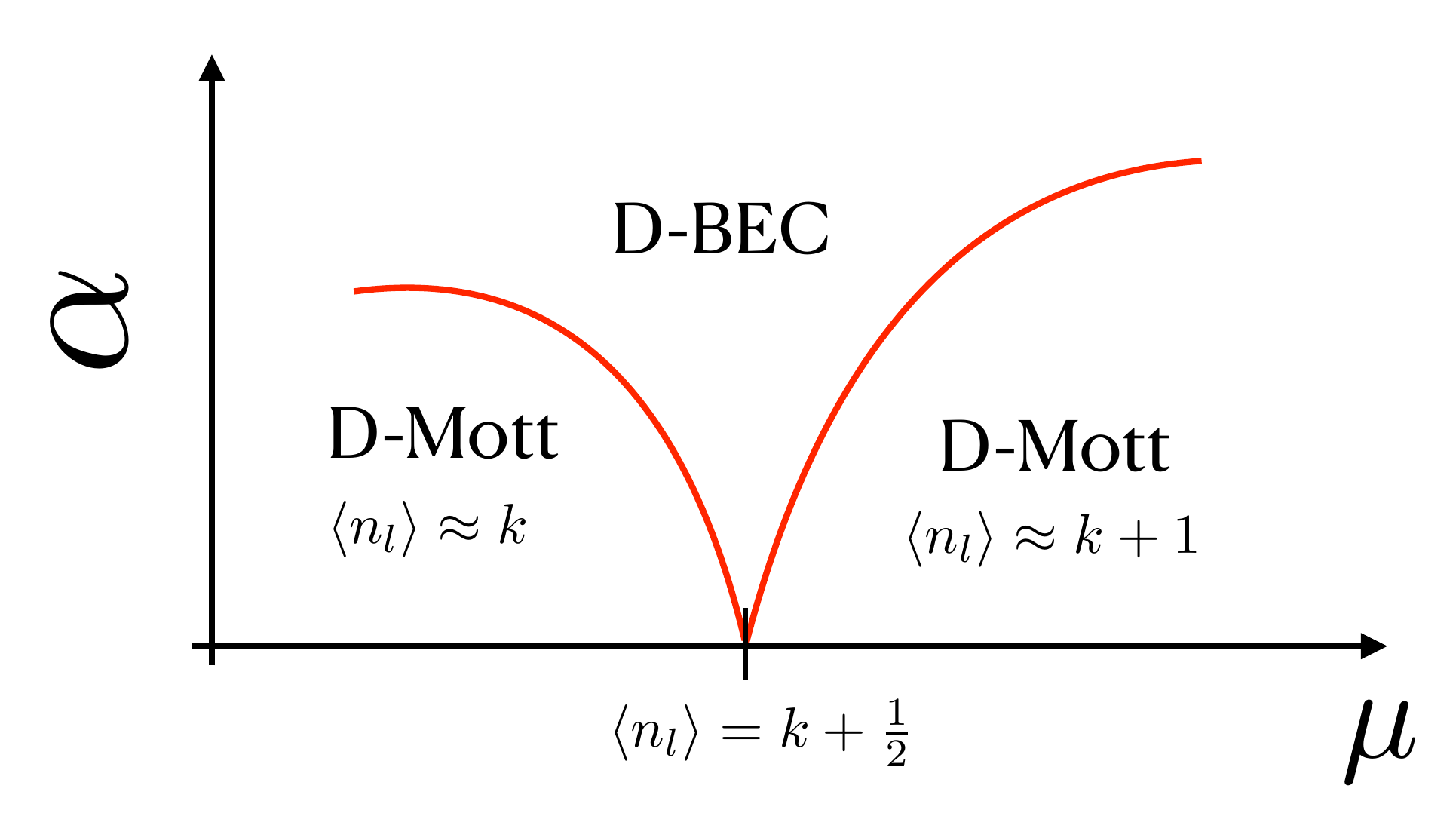}
    \caption{Sketch of the D-Mott/Metal phase diagram as a function of the chemical potential $\mu$ and the dissipation strength $\alpha$; $k$ is an integer.
    The boson-boson interactions are assumed to be short ranged, meaning that the Luttinger parameter $K$ in Eq.~\eqref{eq:act2} fulfills $K > \tfrac{1}{2}$. When  $n_0$ ($N_0$) is tuned to a half-integer value, the dissipative BEC (D-BEC/Superconductor) phase is the ground state for non-vanishing but non-vanishing but arbitrarily small  $\alpha$. The D-Mott lobes are  compressible (unlike their Mott-insulating counterparts) and therefore the lattice filling is not exactly an integer. However, since they derive from the incompressible Mott lobes at $\alpha = 0$, we label them using  integers $k$ corresponding to  the approximate values of the lattice filling.}
    \label{fig:phasediag2}
\end{figure}

For larger  incommensurability such that $\delta^{-1} \lesssim \xi_u$  the puddle picture breaks down and D-Mott phase is not stable.  In what follows, we shall approach the breakdown of the puddle picture by studying a weakly coupled system of  puddles at half-integer filling, i.e. for small $J/E_C$ and for $|N_0| \to \tfrac{1}{2}$ (recall that, as discussed in Sec.~\ref{sec:dmott},  a half-integer $N_0$ implies a half-integer   $n_0$ in the original model, Eq.~\eqref{eq:action}). In this regime, we argue  that the ground state is the D-BEC phase for non-vanishing but arbitrarily small dissipation strength. To show this, we rely on a mapping of the quantum rotor degree of freedom  to a pseudospin (see e.g. Refs.~\cite{Matveev1991,Pruisken_PhysRevB.81.085428}). 
The pseudospin-$\tfrac{1}{2}$  
describes the charge states of the puddle with $n_l  = 1$ and $n_{l}= 0$, which at $|N_0| = \tfrac{1}{2}$   are exactly degenerate in energy. The pseudospin operators $S^{\pm}_l$ cause transitions between these charge states of the  rotor (puddle).
The coupling to the bath is described as an (anisotropic) Kondo coupling to
$N_c$ fermion channels. In the limit where $N_c\to +\infty$, we recover
the local ohmic bath coupled to $S^{\pm}_{l}$ (see Appendix~\ref{app:largenc} for additional details).   In this representation, a weak Josephson tunneling $J$ between the puddles corresponds to an easy-plane Heisenberg coupling. Thus,  a chain of coupled dissipative rotors with half-integer filling can be mapped onto the following  Hamiltonian:
\begin{align}
H &= H_{M} +H_F + H_J,\label{eq:matveev}\\
H_{M} &= h \sum_{l} S^z_l + \frac{t_B}{\sqrt{N_c}}\sum_{l,\alpha} \left[  S^{+}_l \psi^{\dag}_{l\alpha \downarrow}(0) \psi_{l \alpha \uparrow}(0)  + \mathrm{H.c.}\right],\notag\\
H_{F} &= \sum_{k,\alpha,\sigma,l} \epsilon(k) \psi^{\dag}_{l \alpha\sigma}(k)\psi_{l\alpha\sigma}(k),\notag\\
H_J &= -\frac{J_{\perp}}{2} \sum_{l} \left[ S^{+}_{l} S^{-}_{l+1} + \mathrm{H.c.}  \right]
\end{align}
where $h\sim \mu$ controls the deviation of  lattice filling from the mean half-integer filling condition  (i.e. $h = 0$ for 
$|N_0| = \tfrac{1}{2}$);  the Fermi operator $\psi_{l \alpha\sigma}(0) = \sum_{k} \psi_{\alpha\sigma}(k)/\sqrt{L_c}$, where $L_c$
is the length of the channel (measured in units of its Fermi wavelength); $\sigma = \uparrow,\downarrow$ is the spin index,
$\alpha = 1,\ldots, N_c$ the channel index, and $l= 1, \ldots, L$ the lattice site index. 
The single particle dispersion $\epsilon_k$ 
is gapless and yields a constant density of states $\rho_0$ at the Fermi energy (which corresponds to $\epsilon_k = 0$). In constructing this model, we have  omitted the long-range tail of the boson-boson
interaction, which is described by the following Ising term:
\begin{equation}
H_{I} = \sum_{l,l^{\prime}}V_{l,l^{\prime}} S^z_{l}S^{z}_{l^{\prime}} \label{eq:ising}
\end{equation}
This term can be readily incorporated (see below).
However, for the sake of simplicity, we shall first 
discuss the physics of this model by neglecting it. 

The  model \eqref{eq:matveev} is  an easy-plane Heisenberg chain where each spin is coupled 
to a series of independent fermionic baths. It was studied in Ref.~\cite{Lobos_PhysRevB.86.035455} for $N_c = 1$.   Using bosonization, we can treat it  non-perturbatively in the $N_c\to +\infty$ limit.  The   bosonized forms of the pseudospin operators are $S^{\pm}_l \sim e^{\pm i\theta(x_l)}$, where $\theta(x)$ is the phase field~\cite{Giamarchi_book,GogolinTsvelik1999}. Bosonizing the Hamiltonian and   integrating out the fermion channels coupled  in the  $N_c\to +\infty$ limit (cf. Appendix~\ref{app:largenc}), we 
arrive at the following  action:
\begin{align}
S &= S_B[\theta] + S_0[\theta,\phi] + S_u[\phi] + S_D[\theta],\notag\\
S_B[\theta,\phi]&= \int d\tau dx \left[  \frac{i\delta}{4\pi} \partial_{\tau}\theta + \frac{i}{\pi}\partial_x \phi \partial_{x}\theta\right],\notag \\
S_0\left[\theta,\phi \right] &=   \frac{v}{2\pi}  \int  d\tau dx \, \left[ K
 \left( \partial_{\tau}\theta \right)^2 +  K^{-1}\left( \partial_x \phi\right)^2\right],\notag\\
 S_{u}\left[\phi\right] &= \frac{\tilde{g}^{\prime}_u}{\pi a_p} \int d\tau dx\:  \cos \left( 4\phi + x\delta  \right), \notag\\
S_{D}\left[ \theta\right] &=  -\frac{\alpha}{\pi a_p} \int  d\tau d\tau^{\prime}  dx\: f(\tau-\tau^{\prime}) \cos \Delta \theta(x,\tau,\tau^{\prime}), &\notag \\
&\Delta\theta(x,\tau,\tau^{\prime}) = \theta(x,\tau) - \theta(x,\tau^{\prime}),\notag\\
&f(\tau, \beta \to +\infty) = \frac{1}{\tau^2}, \label{eq:matveev}
\end{align}
where $\delta  = \delta(h) = 4\pi \left( |N_0| - \frac{1}{2}\right)/a_p$, with $|\delta| \ll 1$ for the Matveev mapping
to apply.  We notice the above bosonized model is similar to Eq.~\eqref{eq:act2}  with the important difference of the sine-Gordon term  being $\cos \left( 4 \phi + x\delta\right)$ rather than $\cos \left(2\phi + x \delta\right)$. 

 Considering the case $\delta = 0$ first,  we recall the term $\cos 4\phi$ has a larger scaling dimension that $\cos 2\phi$ and it is only  relevant in the RG sense  for $K \lesssim \tfrac{1}{2}$~\cite{Giamarchi_book,GogolinTsvelik1999}. Realizing the latter  requires long range repulsive interactions~\cite{Giamarchi_book}, which map to the Ising term $H_I$ (cf. Eq.~\ref{eq:ising}).  
If we assume that $H_I$ can be neglected because of a strong screening  by the nearby diffusive metal electrode,  then Eq.~\eqref{eq:matveev} for $\alpha = 0$ is the low-energy effective action of the  XX spin chain. For the latter spin-chain,  the Luttinger parameter $K = 1$, which implies that the sine-Gordon term $\cos 4\phi$  is irrelevant  and can be neglected at low energies.   On the other hand, the ohmic dissipation term 
$S_D[\theta]$ is relevant for $K\gtrsim \tfrac{1}{2}$. Since we assume short-range interactions that yield $K\simeq 1$,  $S_D$  turns out to be strongly relevant and drives the system into the ordered D-BEC phase (at small $\alpha$,  the runaway flow   to strong coupling  is described by Eq.~\eqref{eq:rggd} with $K \simeq 1$). Indeed, when the (renormalized) theory is treated semi-classically  in the  $\alpha\sim 1$ limit,   $S_D$ favors long-range order in the phase field $\theta(x,\tau) \to \theta_0  =$ const. for $|\tau|\to +\infty$. The more sophisticated treatment relying on the SCHA discussed in Sec.~\ref{sec:dbec} leads to the same conclusion. 

Slightly away from  half-integer filling, i.e. for $\delta \neq 0$ ($h\neq 0$) for which $|N_0|$ is not exactly  equal to $\tfrac{1}{2}$, the D-BEC phase remains the ground state. In this case, independently of the range and strength of the interactions,  incommensurability makes 
the sine-Gordon term $\sim \cos (4\phi + \delta x)$ in Eq.~\eqref{eq:matveev} irrelevant at low energies. The only relevant (for $K \gtrsim \tfrac{1}{2}$)
term that remains is $S_D[\theta]$, which again favors the D-BEC phase as the ground state of the system. In this case, there is also the possibility of stabilizing a Tomonaga-Luttinger liquid (TLL) as the ground state, as long as the screening of the boson-boson interactions is sufficiently poor and results in  long-range repulsive
tail  $V_{l,l^{\prime}}$ mapping to the Ising term in Eq.~\eqref{eq:ising}. In such case, the Luttinger parameter  is $K < \tfrac{1}{2}$, which  makes $S_{D}[\theta]$ irrelevant in the RG sense (for infinitesimal $\alpha$) and a TLL phase stable.

In retrospect, we may have argued, starting from the bosonized version of model in Eq.~\eqref{eq:act2} and assuming a large incommensurability $ a_0\delta \sim 1$, that we could have arrived at the same conclusion: For large incommensurability, the $\cos(2\phi + \delta x)$ term is irrelevant at low energies, and the only  relevant perturbation that remains for $K \gtrsim \tfrac{1}{2}$ is the quantum dissipation $S_D$, which drives the system to the 
 D-BEC phase (the TLL phase is stabilized for $K \lesssim \tfrac{1}{2}$ where $S_D$ is irrelevant).  Although it is reassuring to reach the same conclusion from two different starting points and the second route appears to be shorter,  we must  be careful. This is because the shorter route ignores the puddle structure of the D-Mott phase and implies  that the D-BEC phase would also be the ground state at  small $\delta$, which is in clear contradiction with what is obtained starting from the puddle picture near commensurate filling. 

The schematic $\alpha$ vs. $\mu$ phase diagram shown in Fig.~\ref{fig:phasediag2} summarizes the results of this section. When the chemical potential $\mu$ is tuned around  integer lattice filling, the D-Mott/metal phase is favored for small dissipation. For larger dissipation $\alpha$, the system transitions to the  ordered  D-BEC/superconductor phase. The quantum criticality of this transition is described by a dissipative TDGL quantum field theory and belongs to the O$(2)$ Wilson-Fisher universality class~\cite{Sachdev_PhysRevLett.92.237003,Werner_JPSP,Pankov_PhysRevB.69.054426} (cf.  
Sec.~\ref{sec:GL}). When $\mu$ is tuned around half-integer lattice filling, for the most realistic case of 
short-range repulsive interactions, 
the D-BEC/superconductor phase is stabilized for non-vanishing but arbitrarily small dissipation $\alpha$. A transition from  D-Mott to D-BEC should be possible at constant dissipation $\alpha$ as a function of the chemical potential $\mu$. By continuity of the phase boundary, we expect the criticality of this transition  to belong to the O$(2)$ Wilson-Fisher Universality class as well. 

\section{Conclusions and Outlook}\label{sec:concl}

In this work, we have studied the superconductor-metal transition in one dimension (1D) 
from  several different theoretical perspectives, which in many cases produce overlapping results. Our main motivation has been to bridge  two existing but completely different  approaches to the superconducting-metal transition in (quasi-) 1D systems~\cite{Sachdev_PhysRevLett.92.237003,Lobos_PhysRevB.86.035455},
as well as relating to  earlier work~\cite{Larkin_PhysRevLett.86.1869}. To this end, rather than studying a very realistic model,  we have simplified the model  down to its bare bones by considering a interacting Bose system on a 1D lattice coupled to an array of ohmic dissipative baths
representing a nearby diffusive metal electrode~\cite{Cazalilla_PhysRevLett.97.076401}. 
By starting from a bosonized description  of this model and using the weak coupling RG analysis  described in Sec.~\ref{sec:rg}, we have estimated the phase boundary between  a phase in which Mott localization dominates (D-Mott/metal, cf. Fig.~\ref{fig:fig1}) and another phase in which the quantum dissipation dominates (D-BEC, cf. Fig.~\ref{fig:fig1}).
 In the strong-dissipation limit and applying the self-consistent harmonic approximation (SCHA) to the bosonized action, we have found the dissipative Bose-Einstein condensate (D-BEC) phase to exhibit off diagonal long-range order. Recalling that the bosons in the present model correspond to fermions bound in the form  of Cooper pairs, this phase is a \emph{bona fide} superconductor with a finite condensate fraction, rather than a fluctuating 1D superconductor with zero condensate fraction at $T = 0$.

 In the regime of weak dissipation, Mott localization/phase slips dominate at commensurate lattice filling. By analyzing the properties of the D-Mott/metal phase appearing in this regime, we have developed a picture of the latter in terms of  dissipative
boson puddles, which mathematically corresponds to an 1D array of Josephson-coupled 
dissipative rotors (cf. Fig.~\ref{fig:puddles}).  The size of the puddles is set by the characteristic length scale over which phase slips suppress coherent tunneling (cf. Fig.~\ref{fig:puddles}). Taking the \emph{na\"ive} continuum limit of the 1D dissipative rotor array  yields a $1+1$ non-linear sigma model (NL$\sigma$M) which is able to describe the properties of both  D-Mott and D-BEC phases. In addition, it also provides (see Appendix~\ref{app:largeN}) a reasonable estimate of the critical exponents at the quantum phase transition between the two phases. In particular, the dynamical exponent turns out to be $z\approx 2$, which is different from $z = 1$ characterizing the Berezinskii-Kosterlitz-Thouless transition between a (fluctuating) superconductor and Mott insulator in 1D~\cite{GogolinTsvelik1999}. 
Using the puddle picture, we have also studied the effect of doping by means of a perturbation expansion in the dissipation strength and the inter-puddle Josephson tunneling. The latter shows that the puddle array has a  finite compressibility and implies that the D-Mott phase is stable against a small doping. Finally, by using a mapping to a pseudospin which applies to the dissipative rotor near half-integer lattice filling~(see e.g. \cite{Matveev1991,Pruisken_PhysRevB.81.085428}) and combined with bosonization, we have have argued that the ground state  of the system is the  D-BEC phase for non-vanishing but arbitrarily small dissipation strength. A qualitative phase diagram as a function of the dissipation strength $\alpha$ and chemical potential $\mu$ is shown Fig.~\ref{fig:phasediag2}. 

Besides illuminating the origin of puddle structure underlying the D-Mott/metal phase, our results  offer useful insights into the phase diagrams predicted in earlier work~\cite{Cazalilla_PhysRevLett.97.076401,Werner_JPSP}. 
In this regard, we first notice that  the derivation of NL$\sigma$M in~\cite{Cazalilla_PhysRevLett.97.076401} relating  models I and II studied in this article to the classical 2D XY lattice model with anisotropic long-range interactions studied by Werner \emph{et al.}~\cite{Werner_JPSP} was not carefully carried out by  taking the continuum limit of  lattice model. Instead,  the bosonized action of a dissipative TLL (described by Eq.~\ref{eq:act3} with $g_u = 0$) was rewritten as the O$(2)$  NL$\sigma$M and subsequently analyzed in the large-$N$ limit. Because no underlying lattice is assumed,
the physical interpretation of the disordered phase  of the NL$\sigma$M is rather unclear. 
Indeed, as shown in this work, the disordered phase corresponds to an array of weakly coupled dissipative rotors (i.e.  puddles). This construction makes no sense without reference to a lattice that gives rise to phase slips. Therefore, in the absence of lattice or at large incommensurability where lattice effects are irrelevant, the only two  possible phases for the models I and II are 1) a Tomonaga-Luttinger liquid in the regime where quantum dissipation is irrelevant, or 2) a phase with long-range order in the regime where quantum dissipation is relevant. This conclusion is consistent with the  recent results obtained by Majumdar \emph{et al.} in Ref.~\cite{Rosso1_PhysRevB.107.165113}, by  numerically studying the bosonized model of the dissipative Luttinger liquid (i.e. Eq.~\ref{eq:act2} with  $g_u = 0$ and $\delta = 0$). These authors found no evidence of the disordered phase predicted by the NL$\sigma$M as used in Ref.~\cite{Cazalilla_PhysRevLett.97.076401}.

 On the other hand, unlike model II studied in~\cite{Cazalilla_PhysRevLett.97.076401}, the model considered in this work is defined on a lattice. At commensurate filling, it maps to the 2D classical  XY lattice model with anisotropic long-range interactions studied by Werner \emph{et al.}~\cite{Werner_JPSP} using (classical) Monte Carlo. Mapping the quantum 1D model to the 2D classical XY lattice model is, strictly speaking,  only possible at commensurate filling, for which the Berry phase term can be dropped.  Away from commensurability, we must be  careful when relying on such quantum-classical mapping. In this work, we have studied the incommensurate boson lattice  by showing that the compressibility of the 1D array of coupled  puddles is finite even at finite incommensurability. Thus, we have argued that a small doping does not destabilize the D-Mott phase (unlike  the case of a 1D MI). On the other hand, 
at large incommensurability, lattice effects are irrelevant, and the correct description is a Gaussian quantum field theory  with the quantum dissipation as the sole perturbation,
i.e. the dissipative Tomonaga-Luttinger liquid first studied in Ref.~\cite{Cazalilla_PhysRevLett.97.076401} using bosonization and weak-coupling RG. 

Besides classical Monte Carlo methods applied in~\cite{Sachdev_PhysRevLett.92.237003,Werner_JPSP} by mapping the model to a classical spin model, the phase diagram quantum 1D boson model~\eqref{eq:act2} has been recently computed using path integral quantum Monte Carlo with  worm  updates  by Ribeiro \emph{et al.}~\cite{ribeiro2023dissipationinduced}. The phase diagram  obtained in Ref.~\cite{ribeiro2023dissipationinduced}  is consistent with the one described
in this work (cf. Fig.~\ref{fig:fig1}). However,  these authors  concluded that the D-Mott phase (called Mott$^*$  in Ref.~\cite{ribeiro2023dissipationinduced}, see their Fig.~1(b)) exhibits a diverging compressibility. This numerical observation is strongly at odds with the  results of the perturbative analysis described in Sec.~\ref{sec:doping} and certainly calls for further numerical investigation of the properties of the D-Mott phase.

Finally, we comment on the differences and similarities of the physics discussed in this work and  the deconfinement transition of an array of 1D Mott insulators studied in Refs.~\cite{Cazalilla_2006,Ho_PhysRevLett.92.130405}. Mathematically, the RG analysis 
described in Sec.~\ref{sec:rg} displays many similarities, with two  phases described as  strong coupling fixed points of two competing relevant perturbations,  namely the Mott potential (which is present in both problems and favors boson localization), and the dissipation/Josephson tunneling (which favors boson delocalization into a BEC phase with off-diagonal long-range order).  Nevertheless, there are also important differences between the quantum phases of the two systems. The most important one is that, in the 1D Mott-insulator deconfinement, bosons remain  largely localized  in the 1D Mott insulating phase even in the presence of non-zero Josephson coupling. This is because, even after hopping to a neighboring 1D system in the two-dimensional lattice of tubes~\cite{Ho_PhysRevLett.92.130405,Cazalilla_2006}, bosons experience  repulsion from other bosons. Therefore, there is little energy to be gained from partial delocalization and the resulting phase remains incompressible. However, in the  dissipative model studied here, bosons can lower their kinetic energy by hopping to the nearby electrode (i.e. the dissipative bath), where they experience no interactions while diffusing in the metal. This major difference has a significant impact on the properties of the D-Mott phase, which therefore acquires a finite compressibility. 

The work reported here may be extended in several possible directions. One such direction is to account for the effect of disorder  in the 1D boson lattice. It is known that, even in the presence of long wavelength disorder, a 1D boson system with commensurate lattice filling undergoes Anderson localization~\cite{Fisher_PhysRevB.40.546}. It is therefore an  interesting to problem understand whether the  puddle picture investigated here remains  valid as a starting point to understand the physics of the 
Anderson-localized (Bose-glass) phase in the presence of dissipation. Another interesting question is the universality class the quantum phase transition to a potentially dissipation-ordered phase, which may be stabilized for large dissipation strength. Not to mention, the question of whether the properties of such ordered phase are affected or not by  the disorder.   

\section*{Acknowledgments} 

 I am grateful to A. Lobos for carefully reading the manuscript and useful comments. 
 This research  has been been supported by the Spanish Ministry of Science and Innovation MCIN/AEI/10.13039/501100011033 through Grant No. PID2023-148225NB-C32 (SUNRISE).

\appendix

\section{Weak-coupling RG equations}\label{app:scaling}

In the bosonized version of the model, it is convenient to introduce the following (vertex) operators:
\begin{align}
V_{2p\phi}(\vec{r}) &= \frac{e^{2 i p \phi(\vec{r})}}{a^{p^2K}_c}, \\
 V_{\theta}(\vec{r}) &= \frac{e^{i\theta(\vec{r})}}{a^{1/4K}_c},
\end{align}
where the short-hand notation of writing $\vec{r} = (v \tau, x)$ has been introduced.
In the above expression, $a_c = (v\tau_c)$ is cutoff.  The scale-invariance
characteristic of the $S_0[\theta,\phi]$  (cf. Eq.~\ref{eq:act2}) sets in for $|\vec{r}| \gg a_c$. 
For the derivation of the RG equations, we will need the following
operator product expansions (OPE), which are to be understand 
as identities holding inside  expectation values taken over the ground 
state of the Tomonaga-Luttinger liquid (TLL) described by $S_0$ in \eqref{eq:act2}
when $\vec{r}_1 \to \vec{r}_2$:
\begin{align}
V_{2p\phi}(\vec{r}_1) V^*_{2p\phi}(\vec{r}_2) &= \frac{1}{|\vec{r}|^{2K}} \Big[1 + 2 i p\: \vec{r}\cdot \partial_{\vec{R}}\phi(\vec{R})   \notag \\
 &\qquad +  \frac{(2 p)^2 i^2}{2!} \left[ \vec{r}\cdot \partial_{\vec{R}}\phi(\vec{R})  \right]^2+ \cdots \Big], \label{eq:ope1}\\
V_{\theta}(\vec{r}_1) V^*_{\theta}(\vec{r}_2) &= \frac{1}{|\vec{r}|^{1/2K}} \Big[1 +  i \: \vec{r}\cdot \partial_{\vec{R}}\theta(\vec{R})   \notag \\
 &\qquad + \frac{i^2}{2!} \left[ \vec{r}\cdot \partial_{\vec{R}}\theta(\vec{R})  \right]^2+ \cdots \Big]
\label{eq:ope2}
\end{align}
where $\vec{R}  = (\vec{r}_1+\vec{r}_2)/2$ and $\vec{r}  = \vec{r}_1-\vec{r}_2$. These expressions hold within an expectation value taken over the ground state of $S_0$ is to be implicitly understood. The OPE of the vertex operators $V_{\theta}$ with $V_{2p \phi}$
contain a power-law prefactor that is an odd function of $\vec{r}$ and yields a vanishing contribution to the RG equations 
upon angular average over $\vec{r}$. By the same token,  the terms linear in $i\partial_{\vec{r}}\theta$ and $i\partial_{\vec{r}}\phi$ also yield vanishing contributions. 

Using the notation introduced above, we can rewrite the bosonized action $S$ as follows:
\begin{align}
S &= S_0 + S_u + S_d,\\
S_u &= \frac{g_u}{2\pi a^{2- p^2 K}_c} \int d\vec{r}  \left[ V_{2p\phi}(\vec{r}) + \mathrm{c.c.}\right],\\
S_D &= -\frac{\alpha}{2\pi a^{1+s-2/K}_c}  \int\limits_{|\vec{r}_1-\vec{r}_2|> a_c} d\vec{r}_1 d\vec{r}_2 \, \frac{\delta(x_1-x_2)}{|\vec{r}_1-\vec{r}_2|^{2-s}} \notag\\ 
&\qquad\qquad \qquad \times\left[ V_{\theta}(\vec{r}_1) V^*_{\theta}(\vec{r}_2) + \mathrm{c.c.} \right].
\end{align}
Here we have generalized the type Mott potential $g_u$ to one with wave number $2 p \pi n_0$  ($p$ being an integer and $n_0$ being the lattice filling at integer filling) and the dissipation from ohmic to a more general one which corresponds
to sub-ohmic  (super-ohmic) for $s < 0$ ($ s > 0$). The dimensionless couplings $\alpha,g_u$ are assumed to small compared to unity.
The TLL action can be written either in terms of $\theta$ or $\phi$ as follows:
\begin{equation}
S_0 =  \frac{1}{2\pi K}\int d\vec{r} \, \left(\partial_{\vec{r}} \phi \right)^2 = \frac{K}{2\pi}\int d\vec{r} \, \left(\partial_{\vec{r}}
 \theta \right)^2.~\label{eq:s0}
 \end{equation}

To derive the RG equations in the weak coupling regime, we consider the expansion the functional-integral expression of the  partition function $Z(a^{\prime}_c)$ of a model with cut-off $a^{\prime}_c = (1 + \delta \ell)a_c$ ($\delta \ell\ll 1$) in its bosonized form. To second order in the dimensionless couplings $g_{u}, \alpha$ we find the following contributions:
\begin{equation}
O(g_u) = - \frac{g_u(\ell+\delta\ell)}{2\pi \left[(1+\delta  \ell) a_c\right]^{2-p^2 K}} \int d\vec{r}\, \langle V_{2p \phi}(\vec{r}) \rangle +\mathrm{c.c.}
\end{equation}
Equating this result to the $O(g_{u})$ in the expansion of $Z(a_c)$ we obtain the leading order renormalization of the coupling $g_u(\ell)$:
\begin{equation}
g_u(\ell) = \frac{g_u(\ell+\delta \ell)}{\left(1+\delta \ell\right)^{2-p^2K}} \Rightarrow \frac{dg_u}{d\ell} = (2-p^2 K)g_u.
\end{equation}
Next, we  consider $O(g^2_u)$
\begin{align}
O(g^2_u) &= +  \frac{g^2_u(\ell+\delta\ell)}{2!  (2 \pi)^2  \left[(1+\delta  \ell) a_c\right]^{4-2p^2 K}} \notag\\ 
& \times  \int\limits_{|r|>(1+\delta \ell)a_c} d\vec{r}_1 d\vec{r}_2 \left[ \langle V_{2p \phi}(\vec{r}_1) V^{*}_{2p\phi}(\vec{r}_2) \rangle + \mathrm{c.c.} \right]
\end{align}
In order to extract the $O(g^2_u)$ contribution to the RG equations, we need split the integral over $\vec{r}=\vec{r}_1-\vec{r}_2$  as follows:
\begin{align}
 \int\limits_{|\vec{r}|>(1+\delta \ell)a_c} d\vec{r}\, \ldots &=   \int\limits_{|\vec{r}|>a_c} d\vec{r}\,  \ldots \notag\\ 
 &\qquad -  \int\limits_{(1+\delta \ell)a_c > |\vec{r}| > a_c} d\vec{r}\,  \ldots \notag\\ 
 \label{eq:split}
\end{align}
The second term on the right hand side is $O(\delta \ell)$ and therefore contributes to the renormalization of
the coupling constant after the OPE, Eq.~\eqref{eq:ope1} has been employed. Keeping only terms that
are even in $\tau$ and leading in gradients of $\phi$, we arrive at the following expression:
\begin{align}
O(g^2_u \delta \ell) &= -  \frac{2\times (2  i p)^2 g^2_u(\ell+\delta\ell)}{2! (2\pi )^2  \left[(1+\delta  \ell) a_c\right]^{4-2K}}\notag\\
\times &\int\limits_{(1+\delta \ell)a_c > |\vec{r}| > a_c} \frac{d\vec{r}}{|\vec{r}|^{2K}}  \int d\vec{R}\,  \langle\left[ \vec{r}\cdot \partial_{\vec{R}} \phi(\vec{R})\right]^2\rangle  \notag\\
&= \frac{g^2_u(\ell) p^2}{\pi^2 a^{4-2K}_c} \int\limits_{a_c(1+\delta\ell) > r >a_c} \frac{r dr}{r^{2K-2}} \int^{+\pi}_{-\pi} d\varphi \cos^2(\varphi) \notag \\
&\qquad \times \int d\vec{R}\, \langle \left[\partial_{\vec{R}} \phi(\vec{R}) \right]^2\rangle + \cdots\notag\\
&= \frac{g^2_u (\ell) p^2  \delta \ell}{2\pi}\int d\vec{r}\, \langle\left[ \partial_{\vec{r}} \phi(\vec{r})\right]^2\rangle + \cdots
\label{eq:gu2}
\end{align}
In the last equation we have relabelled $\vec{R}\to \vec{r}$.
Upon re-exponentiation, this term leads to the renormalization of the Luttinger parameter, $K$. 
We shall delay its treatment
until we have derived the   contributions to the RG flow that arise from the dissipative coupling to the phase, 
which appear to leading order in the dimensionless coupling $\alpha$:
\begin{align}
O(\alpha) &=  \frac{\alpha(\ell+\delta\ell)}{2\pi \left[(1+\delta\ell)a_c\right]^{1+s-1/2K}}\int_{|\vec{r}|> {(1+\delta \ell)a_c}} d\vec{r}_1 d\vec{r}_2 \notag\\
&\frac{\delta(x_1-x_2)}{|\vec{r}_1-\vec{r}_2|^{2-s}} \left[  \langle V_{\theta}(\vec{r}_1)V^*_{\theta}(\vec{r}_2)\rangle + \mathrm{c.c.}\right]
\end{align}
Using Eq.~\eqref{eq:split}, and comparing the result with the $O(\alpha)$ term of the partition function $Z(a_c)$
we obtain the RG equation for $\alpha$:
\begin{align}
\alpha(\ell) &= \frac{\alpha(\ell+\delta\ell)}{\left[ (1+\delta \ell) a_c\right]^{1+s-1/2K}}\notag\\
\Rightarrow& \qquad \frac{d\alpha}{d\ell} = \left(1+s-\frac{1}{2K}\right) \alpha.
\end{align}
The other contribution to $O(\alpha)$ arises from the $O(\delta \ell)$ term  in 
Eq.~\eqref{eq:split} and the OPE~\eqref{eq:ope2}:
\begin{align}
O(\alpha\delta \ell) &= -\frac{i^2 \alpha(\ell) }{2 \pi a^{1+s-1/2K}_c} \int
\limits_{(1+\delta\ell)a_c>|v \tau|> a_c} \frac{v d\tau}{|v\tau|^{-s+1/2K}}\notag\\
&\qquad\qquad\times \int d\vec{R} \, \langle \left[ \partial_{\tau} \theta(\vec{R})/v \right]^2 \rangle\notag\\
&= \frac{\alpha(\ell)}{\pi} \delta\ell \int d\vec{r} \, \langle \left[ \partial_{\tau} \theta(\vec{r})/v \right]^2 \rangle.
\label{eq:gdres}
\end{align}
It is convenient to rewrite this result in terms of the field $\phi$ and together with the $O(g^2_u\delta \ell)$ they  
can be re-exponentiated and used to compute the corrections to the Luttinger parameters $(K, v)$ in the action 
$S_0$ in  terms of $\phi$ (first expression on the right hand-side of Eq.~\ref{eq:s0}).  This can be achieved
within the operator formalism  by using the equations of motion for $\phi$ derived from $S_0$, which yield:
\begin{equation}
\partial_{\tau}\theta(\vec{r}) =  -\frac{i v}{K} \partial_{x} \phi(\vec{r}).
\end{equation}
This equation holds within the expectation values computed with $S_0$. 
Thus, Eq.~\eqref{eq:gdres} becomes
\begin{equation}
O(\alpha\delta \ell) = -\frac{\alpha\delta \ell}{\pi K^2} \int d\vec{r} \, \langle \left[\partial_{x} \phi(\vec{r}) \right]^2 \rangle
\end{equation}
Hence, we see that  the $O(\alpha\delta \ell)$ contribution corrects the coefficient of $(\partial_{x} \phi)^2$ in $S_0$, i.e. $v/K$. Combining it with the $O(g^2_u \delta \ell)$ correction, we obtain:
\begin{align}
\left( \frac{v}{K} \right)(\ell) &= \left( \frac{v}{K} \right)(\ell+\delta\ell)  + v(\ell) \left( \frac{2 \alpha(\ell)}{K^2(\ell)} - \frac{p^2 g^2_u(\ell)}{v(\ell)} \right)\delta\ell\notag\\
&\Rightarrow \frac{d}{d\ell} \left( \frac{v}{K}\right) = v \left(p^2 g^2_u  -  \frac{2\alpha}{K^2}  \right).\label{eq:vj}
\end{align}
Therefore, whilst the Mott potential $\propto g_u$ tends to increase the ratio $v/K$,  the local phase
dissipation $\alpha$ tends to decrease it.  Let us recall that $K/v$ is proportional compressibility~\cite{Cazalilla_JphysB,Giamarchi_book,Cazalilla_RevModPhys.83.1405} and its
tendency to be renormalized to smaller values by the Mott potential is expected, becausethis term dominating  at low-energies results in an
incompressible insulating phase (the 1D Mott insulator). On the other hand, the local phase dissipation phase favors ordering the phase, which tends to increase the compressibility of the TLL phase by further delocalizing of the bosons and favoring a Bose-Einstein condensate.

 Finally, we study how the coefficient of $\left( \partial_{\tau} \phi\right)^2$, namely $1/(vK)$ is renormalized. This parameter
is only corrected by  $O(g^2_u\delta \ell)$ contribution (see ~\eqref{eq:gu2}), which yields:
\begin{align}
\left(\frac{1}{vK}\right)(\ell +\delta \ell) &= \left( \frac{1}{vK}\right)(\ell) - \frac{g^2_u p^2}{v(\ell)} \delta\ell \notag\\
&\Rightarrow \frac{d}{d\ell} \left( \frac{1}{vK} \right) = \frac{p^2 g^2_u }{v}.\label{eq:vn}
\end{align}
From Eq.~\eqref{eq:vj} and \eqref{eq:vn}, we can obtain the RG equations for the Luttinger parameter $K$ 
and the plasmon velocity $v$:
\begin{align}
\frac{d}{d\ell} K^2 &= \frac{d}{d\ell} \left( \frac{1}{1/(v K)} \times \frac{1}{v/K} \right)\notag\\
 &= -K v^3 \frac{d}{d\ell}\left[(v K)^{-1} \right]  -  \frac{K^3}{v} \frac{d}{d\ell} \left( \frac{v}{K}\right) \notag\\
 & = - K v^2  p^2  g^2_u - K^3 \left(p^2 g^2_u - \frac{2\alpha}{K^2} \right)\\
 &\Rightarrow \frac{d K}{d\ell} = \alpha - p^2 K^2 g^2_u. \label{eq:K}\\
 \frac{d}{d\ell} v^2 &= \frac{d}{d\ell} \left( \frac{v}{K} \times \frac{1}{1/(v K)} \right) \notag\\
 &= v K \frac{d}{d\ell} \left(\frac{v}{K} \right) - ( v K)^2  \frac{d}{d\ell} \left( \frac{1}{v K}\right) \notag \\
 &= v^2 K \left( p^2 g^2_u - \frac{2\alpha}{K^2}\right) - v K^2 p^2 g^2_u\\
 & \Rightarrow \frac{d}{d\ell} v = -\frac{v}{K} \alpha.
\end{align}
\section{Large number of channels}\label{app:largenc}

In this Appendix, we study how an ohmic dissipative coupling emerges in the limit
where the number of fermionic channels $N_c$ coupled to a bosonic variable is taken to 
infinity. We consider the following simplified model:
\begin{align}
S[\bar{\psi}, \psi,\bar{\phi},\phi] &= S_{B}\left[ \bar{\phi}, \phi \right] + S_{F}[\bar{\psi},\psi]\notag + S_{K}[\bar{\psi}, \psi,\bar{\phi},\phi]\\
S_F &=   \sum_{\alpha=1}^{N_C} \sum_{k,\sigma}\int d\tau \, \bar{\psi}_{\alpha\sigma}(k)\left[ \partial_{\tau} + \epsilon_{k} \right] \psi_{\sigma,\alpha}(k) 
\notag\\
S_{B}&= \frac{t_B}{\sqrt{N}_c}\sum_{\alpha=1}^{N_c} \int d\tau \left[  \phi s^{+}_{\alpha} + s^{-}_{\alpha} \phi \right],
\end{align}
where $s^{+}_{\alpha} = \bar{\psi}_{\alpha\uparrow}(0) \psi_{\alpha\downarrow}$ and $s^{-}_{\alpha}(0) = \bar{\psi}_{\alpha\downarrow}(0) \psi_{\alpha\uparrow}(0)$, where $\psi_{\alpha\sigma}(0) = \sum_{k} \psi_{\alpha\sigma}(k)/\sqrt{L_c}$, where $L_c$ is the length of the channel in units of the channel's Fermi wavelength. $S_B[\bar{\phi},\phi]$ describes the dynamics of the local bosonic
variables $\bar{\phi}$ and $\phi$, and since it will not be needed for our discussion below, we shall omit writing its
explicit form.

In order to make the model well defined  in the $N_c\to +\infty$ limit, we have introduced a coupling  that is multiplied by a factor of $1/\sqrt{N_c}$. The $N_c$ non-interacting spin-$\tfrac{1}{2}$ fermion channels are described by $S_F[\bar{\psi},\psi]$, where $\psi_{\alpha\sigma}(k)$ are Grassmannian fields. The (tunneling) density of states of all channels at the Fermi energy is assumed to be a constant, $\rho_0$. Thus,  the (zero-temperature) two-point correlators
\begin{equation}
g_0(\tau) = - \langle \psi_{\alpha,\sigma}(\tau) \bar{\psi}_{\alpha^{\prime},\sigma^{\prime}}(0)\rangle 
\end{equation}
asymptotically behave as $g_0(\tau \gg \tau_c) \simeq \rho_0/\tau$, for $\tau_c \sim D^{-1}$, where $D$ is the channel
bandwidth. The result of integrating out the fermion channels is the following ``influence functional''
\begin{align}
\mathcal{F}[\bar{\phi},\phi] = \int \left[ d\bar{\psi} d\psi\right] e^{-S_F[\bar{\psi},\psi]- S_B[\bar{\psi},\psi,\bar{\phi},\phi]}.
\end{align}
Using the cumulant expansion, we find
\begin{multline}
S_D[\bar{\phi},\phi] = \log \mathcal{F}[\bar{\phi},\phi]\\ 
= \sum_{n=0}^{+\infty}  \frac{(- g^{2}_B)^n}{N^{n}_c}
\int d\tau_1 \dots d\tau_{2n}\,   \bar{\phi}(\tau_1)\cdots \bar{\phi}(\tau_n)\\
\times C^{(2n)}_s(\{\tau_i\}_{i=1}^n|\{\tau_{j}\}_{j=n+1}^{2n})\phi(\tau_{n+1}) \cdots \phi(\tau_{2n}),
\end{multline}
where 
\begin{align}
 C^{(2n)}_s(\{\tau_i\}_{i=1}^n|\{\tau_{j}\}_{j=n+1}^{2n}) &= (-1)^n \langle S^{+}(\tau_1) \dots S^{+}(\tau_n) \notag\\
 \times &S^{-}(\tau_{n+1})\cdots S^{-}(\tau_{2n})\rangle_c
\end{align}
is the fully connected correlator of  $S^{\pm}(\tau) = \sum_{\alpha=1}^{N_c} s^{\pm}_{\alpha}(\tau)$,
where $\langle \ldots \rangle_c$ stands for the cumulant average that contains only fully connected contractions
of the fermion correlators. Since the fermion channels are independent, it can be seen that $G_c = O(N_c)$. 
For example, explicit evaluation yields  ($\tau_{ij} = \tau_{i}-\tau_{j}$):
\begin{align}
C^{(2)}_s(\tau_1,\tau_2) &= -N_c g_0(\tau_{12}) g_0(\tau_{21})\notag\\
                                  &\sim \frac{N_c \rho^2_0}{(\tau_{12})^2},\\
C^{(4)}_s(\tau_1,\ldots,\tau_4) &= N_c  \left[ \det M(\tau_1,\ldots \tau_4) \right]^2
\end{align}
where the $2\times 2$ matrix
\begin{equation}
M(\tau_1,\ldots,\tau_4) = \left( 
\begin{array}{cc}
g_0(\tau_{13}) & g_0(\tau_{14})\\
g_0(\tau_{23}) & g_0(\tau_{24})
\end{array}
 \right).
\end{equation}
While the $O(g^2_B)$ term is $O(1)$, the $O(g^4_B)$ is $O(1/N_c)$. Higher order
terms are suppressed by higher powers of $N_c$, with the $O(g^{2n}_B)$ term being $O(1/N^{n-1}_c)$. In the $N_c\to +\infty$ limit, only the $O(g^2_B)$ survives and we obtain:
\begin{equation}
S_D[\bar{\phi},\phi] = -t^2_B \rho^2_0 \int\limits_{|\tau-\tau^{\prime}|> \tau_c} d\tau d\tau^{\prime}\: \frac{\bar{\phi}(\tau)  \phi(\tau^{\prime})}{|\tau-\tau^{\prime}|^2}.\label{eq:largeNc}
\end{equation}
Finally, let us notice that here and  in the main text we have assumed the bosonic variables. $\phi,\bar{\phi}$ couple to the spin
fluctuations of the fermion channels through the operators $s^{\pm}_{\alpha}$. However,  
coupling to the superconducting fluctuations of the non-interacting channels via:
\begin{equation}
S_{B}= \frac{t_B}{\sqrt{N}_c}\sum_{\alpha=1}^{N_c} \int d\tau \left[  \phi \bar{\Delta}_{\alpha} + \Delta_{\alpha} \bar{\phi} \right],
\end{equation}
where $\Delta_{\alpha}= \psi_{\alpha\downarrow}(0) \psi_{\alpha \uparrow}(0)$ and 
 $\Delta_{\alpha}= \bar{\psi}_{\alpha\uparrow}(0) \bar{\psi}_{\alpha \downarrow}(0)$ yields the same result (i.e. an Ohmic bath,  Eq.~\ref{eq:largeNc}) in the limit $N_c\to +\infty$. 

\section{The NL$\sigma$M at large $N$}\label{app:largeN}

 In this Appendix, we review the most important results concerning the correlations 
of the dissipative rotor model. It is convenient to rewrite the action for the 
latter in terms of a two-dimensional  vector $\vec{n}(\tau) = \left( \cos \theta(\tau), \sin \theta(\tau) \right)$
which unit length, i.e. $\vec{n}^2(\tau) = 1$ for $0<\tau < \beta = 1/T$. Using this constraint and  $\int d\tau F (\tau) =0$ (where $f(\tau) = \beta^{-1}\sum_{\omega_m} |\omega_m| e^{-i\omega_m \tau} = \tau^{-2}$ in the $\beta \to +\infty$ limit), we can rewrite the $0+1$ dissipative rotor model as follows:
\begin{align}
S_D[\vec{n},\lambda] &= \frac{\alpha_p}{4} \int d\tau d\tau^{\prime} f(\tau-\tau^{\prime}) \left[ \vec{n}(\tau) - \vec{n}(\tau^{\prime})\right]^2 \notag\\
& + \frac{1}{2E_C}\int d\tau \left( \partial_{\tau} \vec{n}\right)^2 + \frac{i}{2} \int d\tau \lambda \left( \vec{n}^2 - 1 \right)
\end{align}
The Lagrange multiplier $\lambda(\tau)$ is introduced to enforce the unit length constraint of 
$\vec{n}(\tau)$. Expanding $\vec{n}(\tau) = \sum_{\omega_m} e^{-i\omega_m \tau} \vec{n}(\omega_m)/\beta$ (with $\omega_m = 2\pi m/\beta$),
the $\vec{n}$-dependent part of $S_D[\vec{n},\lambda]$ can be written as
\begin{align}
S_0[\vec{n}] &= \frac{1}{2\beta}\sum_{\omega_m} G^{-1}(\omega_m)
|\vec{n}(\omega_m) |^2,\\
G^{-1}(\omega_m) &=   \alpha |\omega_m|  +  \frac{\omega^2_m}{E_C}.\label{eq:svecn}
\end{align}
In order to analytically study this model, we shall generalize the symmetry group of this model  from $O(2)$ to $O(N)$ symmetry, which amounts to assuming that  
$\vec{n} = \left(n_1, \ldots, n_{N}\right)$ with $N$ arbitrarily large. Furthermore, 
since the action for $\vec{n}(\tau)$ is Gaussian, we can proceed to integrate this vector field out, which results in the following expression:
\begin{equation}
S[\lambda] =     \frac{N}{2} \mathrm{Tr} \log \left[ G^{-1} + i \lambda \right] - \frac{i}{2}   \int d\tau  \lambda(\tau).
\end{equation}
In the large-$N$ limit, $S[\lambda]\sim O(N)$ and the functional integral for the partition function $Z = \int \left[ d \lambda \right] \, e^{-S[\lambda]}$ 
is dominated by the saddle point of the action,  which is obtained from the condition
\begin{align}
&\frac{\delta}{\delta \lambda(\tau)} S[\lambda] = 0 
 \Rightarrow \frac{N}{2}  \mathrm{Tr}   \left[ \frac{1}{G^{-1} + i \lambda} \right] = 1.\label{eq:saddle}
\end{align}
For $N\gg 1$, the saddle point is dominated by constant solutions of the form 
$\lambda_0 = -i \alpha_p \Delta$ with $\Delta  > 0$. As shown below, such constant solution yields
exponentially decaying correlations, i.e. a disordered phase. Note that $\lambda = 0$
(or, for that matter $\lambda = -i \alpha_p \Delta$ with $\Delta < 0$) are not acceptable solutions 
because the correlation
function at the saddle point would be singular for ohmic dissipation  (the presence of a 
$\lambda = 0$ solution implies the  existence of an ordered state, which is not possible for ohmic dissipation but it is possible for sub-ohmic dissipation~\cite{Renn1,Renn2}). For $\lambda$ real, the correlations
would be oscillatory in $\tau$ which is also not physically acceptable for a dissipative system.
By expanding  the action, around the constant saddle point solution, i.e. setting 
$\lambda = \lambda_0 +  u(\tau)$, where $\int d\tau u(\tau) = 0$, we see
the fluctuations are controlled by
\begin{align}
\frac{1}{2}\int d\tau d\tau^{\prime} u(\tau) \frac{\delta^2 S[\lambda]}{\delta \lambda(\tau) \delta \lambda(\tau^{\prime)}} u(\tau^{\prime}) \notag\\
\propto \frac{N}{4\beta} \sum_{\omega_m} \frac{|\delta u(\omega_m)|^2}{\left( G^{-1}(\omega_m) + \alpha_p \Delta \right)^2}, 
\end{align}
which is $O(N)$ and therefore will suppress any fluctuations of $\lambda(\tau)$ around $\lambda_0$
in the $N\to +\infty$ limit. Thus, Eq.~\eqref{eq:saddle} becomes:
\begin{equation}
\frac{N\beta}{2} \sum_{\omega_m} \frac{1}{\alpha|\omega_m| + \omega^2_m/E_C + \alpha_p \Delta} = 1.\label{eq:condN}
\end{equation}
At zero temperature where $\beta\to +\infty$, the solution of this equation is aided
by regarding the $\omega^2_m/E_C$ as imposing an effective large frequency cut-off
which is estimated from the condition $\omega^2_C/E_C \approx \alpha \omega_C$, i.e.
$\omega_C \approx E_C \alpha$. Thus,  Eq.~\eqref{eq:condN} becomes:
\begin{equation}
\frac{N}{\alpha_p} \int^{\omega_C}_0 \frac{d\omega}{\pi} \frac{1}{\omega +\Delta} \simeq  \frac{N}{\alpha \pi}\log\left( \frac{\omega_C}{\Delta}\right) = 1, 
\end{equation}
with logarithmic accuracy.
 Hence,
\begin{equation}
\Delta = \omega_C e^{-\pi \alpha_p/ N}.
\end{equation}
Returning to \eqref{eq:svecn}, and evaluating it at the saddle point, we see that the $\vec{n}$ correlations 
become gaussian:
\begin{equation}
\langle \vec{n}(\omega_m) \cdot \vec{n}(-\omega_m)\rangle \simeq \frac{N/\alpha_p}{|\omega_m| + \Delta}
\end{equation}
for $\omega_m\ll \omega_C$. Hence, at $T=0$ using a Gaussian cut-off for simplicity, we obtain:
\begin{align}
G_{\theta}(\tau) &= 2 \langle  \vec{n}(\tau) \cdot \vec{n}(0)\rangle  = \frac{2N}{\alpha_p} \int \frac{d\omega_m}{2\pi} e^{-\omega^2/\omega^2_C}\, \frac{e^{-i\omega_m\tau}}{|\omega_m|+\Delta}\notag\\
&= -\frac{2 N/\alpha_p}{\tau^2} \int  \frac{d\omega_m}{2\pi}e^{-\omega^2/\omega^2_C}\,\left[  \frac{ \partial^2_{\omega_m} e^{-i\omega_m\tau}}{|\omega_m|+\Delta}\right]\notag\\
&=  -\frac{2N/\alpha_p}{ \tau^2} \int  \frac{d\omega_m}{2\pi} \partial^2_{\omega_m}  \left[\frac{ e^{-\omega^2/\omega^2_C}}{|\omega_m|+\Delta} \right] e^{-i\omega_m\tau}\notag \\
&\qquad \sim \frac{1}{\tau^2}.\label{eq:gtheta}
\end{align}
The factor of $2$ in
the first line of \eqref{eq:gtheta}
introduced so that $G_{\theta}(\tau)$ agrees with the definition given in Eq.~\eqref{eq:spohn} for $N =2$.
In order to obtain last expression in \eqref{eq:gtheta} we have used $\partial^{2}_{\omega_m} |\omega_m| = 2 \delta(\omega_m)$.
The result is valid asymptotically for $|\tau| \gg \Delta^{-1} > \omega^{-1}_C\gtrsim \omega_c  = \tau^{-1}_c$,
and agrees (ignoring the constant prefactor) with the exact result for $1\le N \le 4$~\cite{Spohn1999}.

  Next, we turn our attention to the analysis of the  NL$\sigma$M in $1+1$ dimensions introduced as the continuum limit of a 1D array of dissipative rotors discussed in Sec.~\ref{sec:dmott}, Eq.~\eqref{eq:cNLsM}. Introducing
the Lagrange multiplier field that ensures the normalization $\vec{n}(x,\tau) = 1$ at 
every point $(x,\tau)$ the action reads:
\begin{multline}
S[\vec{n}] = \frac{1}{2\beta L} \sum_{q,\omega_m} G^{-1}(q,\omega_m) |\vec{n}(q,\omega_m)|^2\\
+ \frac{1}{2}\int dx d\tau \lambda(x,\tau) \left[  \vec{n}^2(x,\tau) -1\right]
\end{multline}
where $G^{-1}(q,\omega_m) = \eta |\omega_m| + \omega^2_m/\gamma + \kappa q^2$, with $\eta \sim \alpha_p/a_p$, $\gamma \sim a_p E_C$ and $\kappa \sim Ja_p$ ($J$ being the Josephson coupling of
the puddles and $a_p$ lattice parameter of the array of puddles).  Next, we integrate out $\vec{n}(x,\tau)$ after generalizing the symmetry group of the model from O$(2)$ to O$(N)$ which means that $\vec{n}(x,t) = \left(n_1(x,\tau), \ldots,
n_N(x,\tau)\right)$. 
The effective action for $\lambda(x,\tau)$ reads
\begin{equation}
S[\lambda] =  \frac{N}{2} 
\: {\rm Tr} \: \log \left[ G^{-1} + i\lambda \right] - \frac{i}{2} \int dx d\tau\, \lambda(x,\tau), 
\end{equation}
In the large-$N$ limit, the above path integral can be estimated using
the saddle point approximation.  In this case too, the saddle point 
is found by extremizing $S[\lambda]$ and setting $\lambda(x,\tau) = \lambda_0 = {\rm const.}$, which leads to
\begin{equation}
 \frac{N}{\beta} \sum_{i\omega_m}\int \frac{dq}{2\pi} \frac{1}{\eta |\omega_m| + \kappa q^2 + i\lambda_0} = 1.
\end{equation}
Note that, in the expression above, the term of $O(\omega^2_m)$ has been neglected because the low frequency behavior is dominated by the dissipative term $\sim \eta |\omega_m|$.
As in the case  of the dissipative rotor in $0+1$ dimensions, the saddle point is located in the imaginary axis, i.e. at $\lambda_0 = -i \kappa \xi^{-2}_c$. Thus, 
at zero temperature, the saddle point condition becomes:
\begin{equation}
N\int \frac{dqd\omega_m}{(2\pi)^2} \frac{1}{\eta |\omega_m| +
 \kappa (q^2 + \xi^{-2}_c)} = 1.
\end{equation}
If we introduce $\bar{\omega} = \eta \omega_m$ and  $\bar{q} = \kappa^{1/2}
q$,  and $g = (\kappa \eta^2)^{-1/2}$, we can rewrite the above expression as follows: 
\begin{equation}
gN  \int \frac{d\bar{\omega} d\bar{q}}{(2\pi)^2} \, \frac{1}{|\bar{\omega}|
+ \bar{q}^2 + \kappa \xi^{-2}_c} = 1.\label{largeN} 
\end{equation}
At the critical point where $g = g^*$, the correlation length diverges, i.e. 
$\xi^{-1}_c = 0$, which yields
\begin{equation}
g^* N  \int \frac{d\bar{\omega} d\bar{q}}{(2\pi)^2} \, \frac{1}{|\bar{\omega}|
+ \bar{q}^2} = 1 
\end{equation}
Upon adding and subtracting this expression,  Eq.~\eqref{largeN}, can be rewritten as
\begin{align}
1&= gN  \int \frac{d\bar{\omega} d\bar{q}}{(2\pi)^2} \, \left[ \frac{1}{|\bar{\omega}|
+ \bar{q}^2 + \kappa \xi^{-2}_c}   - \frac{1}{|\bar{\omega}|
 + \bar{q}^2} \right] \notag\\
&\qquad + g N \int \frac{d\bar{\omega} d\bar{q}}{(2\pi)^2} \,
 \frac{1}{|\bar{\omega}| + \bar{q}^2}  \notag \\ 
&=  - gN \int \frac{d\bar{\omega} d\bar{q}}{(2\pi)^2} \,
\frac{\xi^{-2}}{(|\bar{\omega}| + \bar{q}^2)(|\bar{\omega}| + \bar{q}^2
+ \kappa \xi^{-2}_c)} \notag\\
&\qquad + \frac{g}{g^*}.
\end{align}
The integral on the right-hand side of the last expression above is convergent for
$\bar{q}$ and $\bar{\omega}$ large and  can be
evaluated  to yield:
\begin{equation}
1  = - g N \frac{\kappa^{1/2} \xi^{-1}_c}{\pi} + \frac{g}{g^*}.
\end{equation}
Hence (recall that $g \propto \eta^{-1} \sim a_p/\alpha_p$)
\begin{equation}
\xi^{-1}_c = \frac{\pi}{N \kappa^{1/2}} \left( \frac{1}{g^*} - \frac{1}{g}\right)\sim
\left( \eta-\eta^*\right)^{\nu} \label{eq:xic}
\end{equation}
for $\eta  < \eta^*$  with $\nu = 1$.  
Thus, at the critical point $\eta  = \eta^*$ the
correlator of $\vec{n}(q,\omega)$ has no characteristic scale:
\begin{equation}
\langle \vec{n}^*(q,\omega_m) \cdot \vec{n}(q,\omega_m) \rangle = \frac{ (\hbar \beta L) N}{\eta |\omega_m| +
\kappa q^2}.
\end{equation}
From this expression the dynamical exponent can be read off, $z = 2$.
Indeed, the values of  two exponents  $z$ and $\nu$ are fairly close to those obtained in Refs.~\cite{Pankov_PhysRevB.69.054426,Sachdev_PhysRevLett.92.237003}  using an $\epsilon$-expansion for the O($N$) Hertz-Moriya-Millis theory:
\begin{align}
z &= 2 - \epsilon^2 \frac{(N+2)(12-\pi^2)}{4(N+8)^2}  + O(\epsilon^3), \\
\nu &= \frac{1}{2} +\epsilon \frac{(N+2)}{4(N+8)}  + \epsilon^2 (N+2)\times\notag \\
&\quad\frac{\left[N^2+N(38-\frac{7\pi^2}{6}) + 132 - \frac{19\pi^2}{3}\right]}{8 (N+8)^3}  +
O(\epsilon^3).
\end{align}
Letting  and $\epsilon = 2-d = 1$ and $N\to +\infty$,  we see that $z \to 2$ and $\nu \to 7/8$ ( $\nu =1.97(3)$ and $\nu = 0.689(6)$ are the exponents computed using Monte Marlo~\cite{Werner_JPSP,Sachdev_PhysRevLett.92.237003} for $N = 2$).

The value of the critical coupling $\eta^* \sim \alpha^*_p/a_p$ is non-universal, but can be estimated
by restoring the $\omega^2_m$ dependence in $G(q,\omega_m)$ (cf. Eq.~\ref{eq:cNLsM}). Thus, 
the saddle point condition becomes:
\begin{equation}
N \int \frac{dqd\omega}{(2\pi)^2} \frac{1}{\eta |\omega| +  \omega^2/\gamma
+ \kappa \left( q^2 +  \xi^{-2}_c\right) ]} = 1.
\end{equation}
Performing the integral over $q$  using Cauchy's theorem yields and setting $\eta = \eta^*$ (and
therefore $\xi^{-1}_c = 0$, cf. Eq.~\ref{eq:xic}), we arrive at 
\begin{align}
1&=  \frac{N}{2\pi}\sqrt{\frac{\gamma}{\kappa}} \log \left[ \frac{4 \omega_c }{\eta^* \kappa} \right]
\end{align}
with logarithmic accuracy.  Hence, since $\xi^{-1}_{c} = 0$ for $\eta = \eta^*$, we obtain:
\begin{equation}
\eta^* \simeq  \frac{4 \omega_c}{\kappa} \: e^{- \frac{2\pi}{N}\sqrt{\frac{\kappa}{\gamma}}}. 
\end{equation}

Finally,  the $\vec{n}$-correlations at the saddle-point can be obtained":
\begin{align}
G_{\theta}(x,\tau) &\propto \langle \vec{n}(x,\tau) \cdot \vec{n}(0,0) \rangle \notag\\
&= N  G[\lambda = -i \kappa \xi^{-2}_c](x,\tau) \notag\\
&= N \int \frac{dqd\omega}{(2\pi)^2} \, \frac{e^{i(qx  - \omega \tau)}}{\eta  |\omega| + \kappa (q^2 + \xi^{-2}_c)}
\end{align}
The equal-time correlation function can be obtained by setting $\tau =0$ and
carrying out the integral over $q$:
\begin{align}
G_{\theta}(x,0) \propto \frac{N}{\kappa^{1/2}} \int^{\infty}_{0} \frac{d\omega}{2\pi} \, 
\frac{e^{- \kappa^{-1/2}|x|\sqrt{\eta \omega + \kappa \xi^{-2}_c } }}{\sqrt{\eta \omega + \kappa \xi^{-2}_c}}.
\end{align}
Performing a change of variables where $u = \kappa^{-1/2 }\sqrt{\eta \omega + \kappa \xi^{-2}_c}$ leads to
\begin{equation}
G_{\theta}(x,0) \propto \frac{N}{\pi \xi_c} \int^{+\infty}_{\xi^{-1}_c} du \: e^{- |x| u} = 
 \frac{N}{\pi \eta  |x|} e^{- |x|/\xi_c},\label{largex}
\end{equation}
which shows that $\xi_c$ is indeed the (renormalized) correlation length of the system in the D-Mott phase. 

Computing local phase correlations can be done with the help of an exponential high-frequency cut-off:
\begin{align}
G_{\theta}(0,\tau) &\propto  \langle \vec{n}(x,\tau) \cdot \vec{n}(0,0) \rangle \notag\\
&= N \int \frac{dqd\omega}{(2\pi)^2} \, \frac{e^{-i \omega \tau}}{\eta  |\omega| + \kappa (q^2 + \xi^{-2}_c)} \notag \\
&=  N {\rm Re} \left[  \int^{+\infty}_0 \frac{d\omega}{2\pi \kappa}  \frac{e^{-\omega ( \tau_c + i\tau)}}{\sqrt{\xi^{-2}_c + \eta \omega / \kappa}} \right]  \notag\\
 &= \frac{N}{\pi} {\rm Re} \left[ e^{(\tau_c + i \tau) \kappa/(\eta \xi^2_c)} 
 \int^{\infty}_{\xi^{-1}_c} du \, e^{- \kappa u^2 (\tau_c + i \tau)/\eta} \right] \notag \\
 &= A(\tau_c) \frac{N}{4\pi} \frac{\eta^2}{\kappa^2 \tau^2} + O(\tau^{-3}),\label{longtau}
\end{align}
where $A(\tau_c)$ is a cut-off dependent prefactor.
In the last expression has been obtained from the  asymptotic expansion 
of the error  function:
\begin{equation}
{\rm Erf}(z) = 1 + \frac{e^{-z^2}}{\sqrt{\pi}} \left[ -\frac{1}{z} + \frac{1}{2z^3
} +  O\left(\frac{1}{z^5}\right) \right]. 
\end{equation}
in the limit where $\tau \gg \tau_c$. 

At the critical point $\eta = \eta^*$ the correlation length diverges, i.e. $\xi^{-1}_c = 0$, and  the critical correlations up to $O(\frac{1}{N})$ corrections can be obtained by setting $\xi^{-1}_c = 0$. Thus, 
\begin{align}
G_{\theta}(x, 0) &\propto \frac{N}{\sqrt{\kappa}} \int^{\infty}_{0} \frac{d\omega}{2\pi} \, 
\frac{e^{- \sqrt{\kappa} |x|\sqrt{\eta^* \omega} }}{\sqrt{\eta^* \omega}}\notag\\
&\sim 
   \frac{N}{\pi \eta^* |x|}.
\end{align}
The local phase correlations  at the critical point can be
obtained using a smooth (e.g. exponential) cutoff:
\begin{align}
G_{\theta}(0,\tau)  &= \int \frac{dq d\omega}{(2\pi)^2} \frac{e^{-i \omega \tau}}{\eta^*  |\omega| + \kappa q^2}\notag\\
&=  \frac{N}{\sqrt{\kappa \eta^*}} {\rm Re} \: \left[ \int^{\infty}_0 \frac{d\omega}{\pi}  \frac{e^{-i\omega\tau} \: e^{-\omega\tau_c}}{\sqrt{\omega}} \right]\notag\\ 
&=  \frac{N}{2\pi \sqrt{\kappa \eta^*}}\:  
{\rm Re} \: \left[ \int^{+\infty}_0  du\,  e^{-u^2(\tau_c + i\tau)} \right] \notag \\ 
&\simeq
\frac{N}{4\sqrt{\pi \kappa \eta^* |\tau|}} + O\left(|\tau|^{-3/2}\right),  
\end{align}
for $\tau \gg \tau_c$. Thus we see that whereas $G_{\theta}(x,0) \sim x^{-1}$ at the 
critical point, $G_{\theta}(0,\tau) \sim \tau^{-1/2}$, which is a consequence of
the dynamical exponent being $z = 2$ in the large-$N$ limit.

 Finally, we considered the correlations in the ordered phase, i.e. for  $\eta > \eta^*$. In order to study this regime in the large $N$ approach to the NL$\sigma$M,
  $n_1(x,\tau) = C_0$, and the remaining $N-1$ components of $\vec{n}(x,\tau)$, i.e.  $\left( n_2(x,\tau), \ldots, n_N(x,\tau)\right)$ are integrated out. Thus, the
 effective action for the Lagrange multiplier becomes:
\begin{multline}
S[\lambda,C_0] =  \frac{i}{2} \int dx d\tau\, \lambda(x,\tau)(C^2_0 -1) \\+ \frac{N-1}{2} 
\: {\rm Tr} \: \log  \left[ G^{-1} + i\lambda\right]
\end{multline}
Again, the large $N$ limit can be studied by 
and extremizing with respect to $\lambda(x,\tau)$ and $C_0$ and 
setting $\lambda(x,\tau) = -i \kappa \xi^{-2}_c$. The following
saddle-point equations are obtained:
\begin{align}
\xi^{-2}_c C_0 &= 0,\notag \\
 (N-1) & \int \frac{dq d\omega}{(2\pi)^2} \: \frac{1}{\eta |\omega| + 
{\kappa} (q^2 + \xi^{-2}_c)} + C^2_0 = 1.
\end{align}
If  $\xi_c \neq 0$ is assumed in the above equations, then $C_0  = 0$ and we recover
the saddle-point condition for the disordered D-Mott phase 
(in the large-$N$ limit, the difference between $N$ and $N-1$ 
is negligible). However, if $C_0 \neq 0$, then $\xi^{-1}_c = 0$ and  find that
\begin{equation}
C^2_0 = 1 -  (N-1) \int \frac{dq d\omega}{(2\pi)^2} \:  \frac{1}{\eta |\omega| + 
\omega^2/\gamma + \kappa q^2)}.
\end{equation}
Note that the integral in the above expression is infrared convergent (but 
ultra-violet divergent, which means that $C_0$ is non-universal) for $\eta \neq 0$. This is consistent with the existence of long range order, as we have assumed. If we repeat the previous steps and derive the correlation function $G_{\theta}(x,\tau) \propto \langle \vec{n}(x,\tau) \cdot \vec{n}(0,0)\rangle$, we  obtain: 
\begin{align}
G_{\theta}(x,\tau) &\propto \langle \vec{n}(x,\tau) \cdot \vec{n}(0,0)\rangle \notag\\
&= C^2_0 + (N-1) H_0(x,\tau),
\end{align}
where 
\begin{equation}
H_{0}(x,\tau) \simeq \int \frac{dqd\omega}{(2\pi)^2} \frac{e^{i(qx - i\omega\tau)}}{\eta |\omega| + \kappa q^2}.
\end{equation}
This function, which describes the contribution of the Goldstone modes, has the same form as $G_{\theta}(x,\tau)$ at the critical point. The asymptotic behavior of the latter has been analyzed above for the critical point correlations and we can borrow those results:
\begin{equation}
G_{\theta}(x,\tau=0) = C^2_0 + (N-1) \frac{\mathcal{A}}{\eta |x|},
\end{equation}
and
\begin{equation}
G_{\theta}(x=0,\tau) = C^2_0 + (N-1) \frac{\mathcal{A^{\prime}}}{\sqrt{\eta \kappa |\tau|}},
\end{equation}
where $\mathcal{A}$ is a cut-off dependent prefactors. 

\section{Perturbative Calculation of the Compressibility}\label{app:pertchi}

In order to compute the compressibility, we shall expand the free energy of the coupled dissipative puddle
array in a double expansion in powers of $\alpha_p$ and $J$. In terms of the partition function of the model, 
the free-energy is:
\begin{equation}
F = - \frac{1}{\beta} \ln Z,
\end{equation}
where the partition function 
\begin{equation}
Z = \int \prod_l \left[ d\theta_l \right] 
e^{-S[\theta_l]},
\end{equation}
where $S[\theta_l]$ has been introduced in  Eqs.~\eqref{eq:act3} and \eqref{eq:act33}. 

We first carry out an expansion in powers of $J$:
\begin{align}
Z &= \int \prod_l \left[ d\theta_l \right] e^{S_P[\theta]} \left( 1 - S_J[\theta_l] + \frac{1}{2!} S^2_J[\theta_l] + \cdots \right) \notag\\ 
&= Z_P \left[ 1 - \langle S_J \rangle_P + \frac{1}{2!} \langle S^2_J \rangle_P + \cdots \right] 
\end{align}
where $S_P[\theta_l] = S_B[\theta_l] + S_{C}[\theta_l] + S_D[\theta_l]$ is the action of a set of independent 
dissipative rotors (puddles) and
\begin{align}
Z_P &= \int \prod_l \left[ d\theta_l \right] \, e^{-S_P[\theta_l]},\\
\langle O \rangle_P  &= \frac{1}{Z_P}  \int \prod_l \left[ d\theta_l \right] \, e^{-S_P[\theta_l]}\: O[\theta_l],\\
\end{align}
Since $S_P[\theta_l]$ is invariant under $\theta_l \to \theta_l \to \theta^{0}_l$, where
$\theta^{(0)}_l$ is a constant that is different for each rotor, the $O(J)$ term vanishes and, as mentioned in the main text, the leading order contribution is $O(J^2)$:
\begin{align}
\frac{1}{2!}\langle S^2_J \rangle_P  &= \frac{J^2}{2!} \sum_{l,l^{\prime}}
 \int^{\beta}_{0} d\tau_1 d\tau_2 \, \langle \mathcal{T} \left\{ \cos\left[ \theta_l(\tau_1) - \theta_{l+1}(\tau_1) \right] \right. \notag\\
 &\left. \qquad \times  \cos \left[\theta_{l^{\prime}}(\tau_2) - \theta_{l^{\prime}+1}(\tau_2)  \right]  \right\} \rangle_P \notag \\
 &= \frac{J^2}{2^2  2!} \sum_{l} \sum_{s=\pm 1} \int^{\beta}_{0} d\tau_1 d\tau_2 \,  \langle \mathcal{T}\left[ e^{i s \theta_{l}(\tau_1)}    e^{- i s\theta_l(\tau_2)} \right] \rangle_P   \notag\\
 &\times\langle \mathcal{T}\left[ e^{-i s \theta_{l+1}(\tau_1)}    e^{ i s\theta_{l+1}(\tau_2)} \right] \rangle_P. \label{eq:2ndorder}
\end{align}
Hence, the free energy:
\begin{align}
F &= -\frac{1}{\beta} \ln Z = F_P - \frac{1}{\beta} \ln \left[ 1 + \frac{1}{2!}\langle S^2_J \rangle_P + \cdots \right] \\
&= F_P - \frac{1}{2! \beta} \langle S^2_J \rangle_P + O(J^4),
\end{align}
where we have used that $\ln (1 + x) = x  + O(x^2)$.  Since the correlation
functions appearing in \eqref{eq:2ndorder} depend on $\tau = \tau_1-\tau_2$, we can
we can change the integration over $\tau_1,\tau_2$ to $\tau$ and $T = (\tau_1+\tau_2)/2$. After taking the limit $\beta\to +\infty$ the following expression for
the leading correction to the ground state is obtained:
\begin{equation}
E_0 - E^{0}_P = -\frac{J^2N_P}{4} \int^{+\infty}_{-\infty} d\tau \, \,
G_{\theta}(\tau) G_{\theta}(-\tau) + O(J^4), \label{eq:corrJ}
\end{equation}
where $N_P$ is the number of puddles (sites) and we have introduced the phase (vertex) correlators:
\begin{equation}
G_{\theta}(\tau) = \langle \mathcal{T} \left[ e^{i\theta(\tau)} e^{-i\theta(0)} \right] \rangle_P.
\end{equation}
In the absence of dissipation (i.e. $\alpha_p = 0$) we can use \eqref{eq:corrJ} 
to show that the correction of the energy is independent of $N_0 = \mu/E_C$ and therefore a weakly Josephson 1D array of quantum rotors at integer filling, which describes the Mott-insulator phase of the system, is incompressible. To this end, we 
use the expression for the two-point phase correlator derived in Appendix~\ref{app:freecorr}, Eq.~\eqref{eq:gtheta2}. Thus we see that the $N_0$ dependence of the ``particle'' propagator $G^0_{\theta}(\tau)$ is exactly canceled
by the ``hole'' propagator $G^0_{\theta}(-\tau)$, which results in~\footnote{In this expression $E^0_P = N_P (E_C N^2_0/2 - E_C N^2_0/2) = 0$ is the ground state energy of
the $N_P$ free rotors (the first term being the contribution from  the charging energy $E_C \sum_l (n_l - N_0)^2/2$ in the ground state where $n_l = 0$ for $l=1,\ldots, N_P$, and the second term is the correction that arises after ``completing 
the square'' and introducing the chemical potential $\mu$ as $N_0 = \mu/E_C$ in the charging
energy).}
\begin{align}
E_0(\alpha_p=0) - E^0_P &=  -\frac{N_P J^2}{4} \int^{+\infty}_{-\infty} d\tau\,
e^{-E_C|\tau|} \notag\\
&= -\frac{N_P J^2}{2 E_C},
\end{align}
i.e. a reduction of the ground state energy due to the hopping which
is independent of $N_0$ and therefore yields a vanishing correction to the
compressibility $\Delta \chi_0 \propto -\partial^2 \left[( E_0(\alpha_p=0) - E^0_P)/N_P\right]/\partial N^2_0$.

In what follows, we turn our attention to the $\alpha_p \neq 0$ case. We first
need to compute the $J = 0$ contribution to $E_P$, i.e. the leading order
correction to the free energy in the decoupled-puddle limit. To this end,
we derive the leading order correction to the ground state energy along
the lines of the expression~\eqref{eq:2ndorder} obtained above. Thus we
obtain:
\begin{equation}
E_P - E^0_P  = -\frac{N_P \alpha_p}{2} \int^{+\infty}_{-\infty} d\tau\, f(\tau) G^{0}_{\theta}(\tau) + O(\alpha^2_p),\label{eq:deltae01}
\end{equation}
where $f(\tau) = \tau^{-2}$ for $\tau\gg \tau_c$.
Rather than the ground-state energy correction, which is dependent on the cutoff $\tau_c$, we are interested in the correction to the compressiblity and therefore, we compute the following integral:
\begin{align}
-\frac{\partial^2 \Delta E^{(1,0)}_0}{\partial N^2_0}  &=  \frac{\alpha_p N_P E^2_C}{2} \int d\tau\, \tau^2 f(\tau) G^{0}_{\theta}(\tau) \notag \\
&\simeq \frac{ N_P \alpha_p  E^2_C}{2} \int d\tau\, e^{N_0E_C\tau} e^{-E_C|\tau|/2} \notag\\
&= \frac{2 N_P \alpha_p E_C}{1-4 N^2_0},    
\end{align}
where $\Delta E^{(1,0)} = \left( E_P - E^0_P\right)$ is given in Eq.~\eqref{eq:deltae01}.
Hence, the $O(\alpha_p)$ correction to the compressibility is 
\begin{align}
\Delta \chi^{(1,0)}_0(N_0) &= -\frac{\partial^2 (\Delta E^{(1,0)}(N_0)/N_P)}{\partial\mu^2}  \notag\\
&=  \frac{\alpha_p}{E_C} \mathcal{C}_1(N_0), \label{eq:chi0alpha}
\end{align}
where we have used the relation $N_0 =  \mu/E_C$ and introduced the function:
\begin{equation}
\mathcal{C}_{1}(N_0) = \frac{1}{2} \int du\,  e^{N_0 u} e^{-|u|/2} = \frac{2}{1-4 N^2_0}.
\end{equation}
At this order (decoupled-puddle
limit), the compressibility of the system is small but finite. Notice also the divergence for $|N_0| = \tfrac{1}{2}$, which is due to  breakdown of
perturbation theory at half-integer filling~\cite{Grabert_PhysRevLett.81.2324,Matveev1991}, where the ground state of the rotor becomes degenerate.

In order to obtain the $O(J^2)$ correction  to the result in Eq.~\eqref{eq:chi0alpha},
we need the $O(\alpha_p)$ correction to the two-point correlator, which can be
obtained from the expression:
(we drop the site index for the $\theta$ field):
\begin{align}
G_{\theta}(\tau) &= \langle \mathcal{T} \left[ e^{i \theta(\tau)} e^{-i \theta(0)}  \right] \rangle_P\\ 
&= \frac{\int \left[ d\theta \right] \,  e^{i \theta(\tau)} e^{-i \theta(0)}    e^{-S_0[\theta] - S_{D}[\theta]}}{\int \left[ d\theta \right] \,    e^{-S_0[\theta] - S_{D}[\theta]} 
}  \notag\\
&=  G^{0}_{\theta}(\tau)  + \frac{\alpha_p}{4}  \sum_{s=\pm}  \int d\tau_1 d\tau_2 \: f(\tau_1-\tau_2) \notag\\
&\times \left[ \langle \mathcal{T}\left[ e^{i\theta(\tau)} e^{is \theta(\tau_1)} e^{-i s\theta(\tau_2)} e^{-i\theta(0)} \right] \rangle_0 \right.\notag \\
&\qquad \qquad \qquad  \left.  -   G^{0}_{\theta}(\tau)  G^{0}_{\theta}(s\tau_1-s\tau_2) \right]  \notag\\
&= G^{0}_{\theta}(\tau)  + \frac{\alpha_p}{2}  \sum_{s=\pm}  \int d\tau_1 d\tau_2 \: f(\tau_1-\tau_2) \notag\\
&\times \left[ \langle \mathcal{T}\left[ e^{i\theta(\tau)} e^{i \theta(\tau_1)} e^{-i \theta(\tau_2)} e^{-i\theta(0)} \right] \rangle_0 \right. \notag\\
&\qquad \qquad \qquad  \left.  -   G^{0}_{\theta}(\tau)  G^{0}_{\theta}(\tau_1-\tau_2) \right].
\end{align}
In the last line we have used that the expression with $s=-1$ can be written as the expression with $s=+1$ if 
we exchange $\tau_1 \leftrightarrow \tau_2$ and use $f(-\tau) = f(\tau)$. Thus the leading order correction is:
\begin{align}
\Delta G^{(1)}_{\theta}(\tau) &= G_{\theta}(\tau) - G^0_{\theta}(\tau)\notag\\ 
&=\frac{\alpha_p}{2} \int d\tau_1 d\tau_2 f(\tau_1-\tau_2) \left[ e^{N_0E_C (\tau+\tau_1-\tau_2)}  \right. \notag\\
&\left.\times e^{E_C(|\tau-\tau_1| - |\tau-\tau_2| - |\tau_1-\tau_2|  - |\tau| - |\tau_1| + |\tau_2|)/2} \right. \notag \\
& \left.   - e^{N_0E_C(\tau + \tau_1 -\tau_2)} e^{E_C(-|\tau| - |\tau_1-\tau_2|)/2} \right].
\end{align}
Hence, using Eq.~\eqref{eq:g4}, we can compute the $O(\alpha_p J^2)$ correction to the ground-state energy:
\begin{align}
&\Delta E^{(1,2)}(N_0) 
= -\frac{ N_P J^2}{2} \int d\tau  \, G^{0}_{\theta}(-\tau)  \Delta G^{(1)}_{\theta}(\tau)\notag \\
&\qquad =  -\frac{N_P\alpha_p J^2}{2}  \int d\tau d\tau_1 d\tau_2  \, f(\tau_1-\tau_2)  \notag \\
&\times\left[ e^{E_C(|\tau-\tau_1| - |\tau-\tau_2| - |\tau_1-\tau_2|  - 2|\tau| - |\tau_1| + |\tau_2|)/2}   \right. \notag\\
 &\left.\qquad  - e^{E_C(-2|\tau| - |\tau_1-\tau_2|)/2} 
\right] e^{N_0 E_C (\tau_1-\tau_2)}.       
\end{align}
Hence, the correction to the second derivative
\begin{align}
-\frac{\partial^2 \Delta E^{(1,2)}}{\partial N^2_0} &\simeq 
\frac{N_P\alpha_p J^2 E^2_C}{2} \int d\tau d\tau_1 d\tau_2 \,  e^{N_0 E_C (\tau_1-\tau_2)} \notag \\
&\times\left[  e^{E_C(|\tau-\tau_1| - |\tau-\tau_2| - |\tau_1-\tau_2|  - 2|\tau| - |\tau_1| + |\tau_2|)/2}   \right. \notag\\
 &\left.\qquad\qquad  - e^{E_C(-2|\tau| - |\tau_1-\tau_2|)/2} 
\right] \notag\\
&= \frac{N_P\alpha_p J^2}{2E_C} \mathcal{C}_2(N_0),
\end{align}
where the function $\mathcal{C}_2(N_0)$ is given by the integral:
\begin{align}
\mathcal{C}_2(N_0) &= 
\int 
 du du_1 du_2  \, e^{N_0 (u_1-u_2)} \notag\\
&\qquad \left[  e^{(|u-u_1| - |u-u_2| - |u_1-u_2|  - 2|u| - |u_1| + |u_2|)/2}  \right.  \notag\\
&\left. \qquad\qquad- e^{(-2|u| - |u_1-u_2|)/2}  \right]  \notag\\
 &= 32  \left[ \frac{15 + 152N_0^2 - 80N_0^4}{(1 - 4N_0^2)^3 (25 - 4N_0^2)}\right].
\end{align}
Hence, the correction to the compressibility is:
\begin{align}
\Delta \chi^{(1,2)}(N_0) &= -\frac{\partial^2 (\Delta E^{(1,2)}(N_0)/N_P)}{\partial \mu^2} \\
&= 
\frac{\alpha_p J^2}{2E^3_C} \mathcal{C}_2(N_0).
\end{align}
Adding this result to the $O(\alpha_p)$ contribution leads to   Eq.~\eqref{eq:compress} 
in the Sec.~\ref{sec:doping}

\section{Correlators of the free rotor model}\label{app:freecorr}

In this section we shall obtain the expression for the phase (vertex) correlators  of the free-rotor model that are used in the perturbative calculation of Appendix~\ref{app:pertchi}.

Let us first consider the action of a free rotor (henceforth we assume  $\beta \to +\infty$):
\begin{align}
S[\theta] &= i N_0 \int d\tau \partial_{\tau} \theta + \frac{1}{2E_C} \int d\tau \, \left( \partial_{\tau}\theta\right)^2 \notag\\ 
&= 
 \frac{1}{2E_C} \int d\tau \left(\partial_{\tau}\theta + i N_0 E_C \right)^2 + \mathrm{const.} \notag\\
  &= \frac{1}{2E_C} \int d\tau \left(\partial_{\tau}\vartheta \right)^2 + \mathrm{const.}
\end{align}
Here we have introduced $\vartheta(\tau) = \theta(\tau) + i N_0 E_C \tau$.
Expanding in Fourier components the new phase field $\vartheta(\tau)$: 
\begin{equation}
\vartheta(\tau) = \int \frac{d\omega_m}{2\pi} e^{-i\omega_m\tau} \vartheta(\omega_m).
\end{equation}
Hence, 
\begin{equation}
S[\vartheta] = \int \frac{d\omega_m}{2\pi}  \frac{\omega^2}{2E_C} |\vartheta(\omega)|^2.
\end{equation}
Thus, the propagator for $\vartheta$ is $E_C/\omega^2$ Let us next compute the  Generating functional for the rotor phase (i.e. vertex) correlation functions:
\begin{align}
Z[J] &=  Z_0 \langle \mathcal{T} \left[ e^{i \int d\tau J(\tau) \theta(\tau)} \right] \rangle_0 \notag\\
&= \int \left[ d\theta \right] \,  e^{-S_{0}[\theta] + i \int d\tau J(\tau) \theta(\tau)}\notag \\
&= e^{N_0 E_C \int d\tau \: \tau J(\tau)}  \notag  \int \left[ d\vartheta \right] \, 
e^{-\frac{1}{2E_C} \int \frac{d\omega_m}{2\pi} \omega^2_m |\vartheta(\omega_m)|^2} \notag \\
&\qquad \times e^{ \frac{i}{2}  \int \frac{d\omega_m}{2\pi}   \left[ J^*(\omega_m) \vartheta(\omega_m) +  \vartheta^*(\omega_m) J(\omega_m)\right]}.
 \end{align}
Upon shifting the fields path integral fields as follows $\vartheta(\omega) \to \vartheta(\omega) + i \frac{E_C}{\omega^2} J(\omega)$ and   $\vartheta^*(\omega) \to \vartheta^*(\omega) + i \frac{E_C}{\omega^2} J^{*}(\omega)$, we arrive at
\begin{align}
Z[J] &= e^{N_0 E_C \int d\tau \: \tau J(\tau)} \left[ \int \left[ d\vartheta \right] e^{-\frac{1}{2E_C} \int \frac{d\omega}{2\pi} |\vartheta(\omega)|^2}  \right] \notag\\
&\qquad\qquad\times \left[ e^{-\frac{E_C}{2}\int \frac{d\omega}{2\pi} \frac{|J(\omega)|^2}{\omega^2} }\right] \\
&= Z_0   e^{N_0 E_C \int d\tau \: \tau J(\tau)}\notag  \\
&\times \exp\left[ -\frac{1}{2} \int d\tau d\tau^{\prime} J(\tau) g_0(\tau-\tau^{\prime}) J(\tau^{\prime}) \right]
\end{align}
where 
\begin{equation}
\tilde{g}_0(\tau) = \int\frac{d\omega}{2\pi}  \left( \frac{E_C}{\omega^2}\right) e^{-i\omega \tau}
\end{equation}
Note that $\tilde{g}_0(\tau)$ is infrared divergent  and needs to be regularized by introducing a low-frequency cut-off in order to be a well defined mathematical object. Nonetheless, this is not a problem for the ``physical'' phase (vertex) correlations  as we will see in what follows. Let us consider a general source for a $n$-point
phase (vertex) correlation function:
\begin{equation}
J(\tau)= \sum_{i=1}^{n} q_i \delta(\tau-\tau_i)
\end{equation}
Hence,
\begin{equation}
Z[J] = Z_0 e^{iN_0 E_c \sum_{i=1}^{n} (q_i \tau_i)} e^{-\frac{1}{2}\sum_{i,j=1}^{n} q_iq_j \tilde{g}_0(\tau_i - \tau_j)}
\end{equation}
Let us rewrite:
\begin{align}
\frac{1}{2}\sum_{i,j=1}^{n} q_i q_j \tilde{g}_0(\tau_i - \tau_j) &= \sum_{i<j=1}^n q_i q_j \tilde{g}_0(\tau_i-\tau_j) \notag\\
&\qquad + \frac{\tilde{g}_0(0)}{2}\sum_{i=1}^{n}  q^2_j. 
\end{align}
The divergence is cured by imposing a neutrality 
condition for the sources, i.e. by requiring that 
\begin{equation}
\int d\tau J(\tau) = \sum_{i=1}^n q_i  = 0.
\end{equation}
Squaring this condition, we can write $\frac{1}{2} \sum_{i=1}^n q^2_i  = -\sum_{i<j=1}^{n} q_i q_j$.
Thus, 
\begin{align}
\frac{1}{2}\sum_{i,j=1}^{n} q_i q_j \tilde{g}_0(\tau_i - \tau_j) &= \sum_{i<j=1}^n q_i q_j \left[ \tilde{g}_0(\tau_i-\tau_j) - \tilde{g}_0(0) \right]  \notag\\
&= \sum_{i<j=1}^{n} q_i q_j g_0(\tau_i-\tau_j) 
\end{align} 
where
\begin{equation}
g_0(\tau) = \int\frac{d\omega}{2\pi}  \left( \frac{E_C}{\omega^2}\right) (e^{-i\omega \tau} - 1) = -\frac{E_C|\tau|}{2}.
\end{equation}
Hence,  
\begin{align}
\langle \mathcal{T} \left[ e^{i\sum_{i=1}^n q_i \delta(\tau-\tau_i)} \right] \rangle_0 &= e^{N_0 E_C \sum_{i=1}^{n} q_i \tau_i} \notag\\
&\times e^{\frac{E_C}{2}\sum_{i< j=1}^{n} q_i q_j |\tau_i - \tau_j|}.
\end{align}
In particular, the two- and four-point vertex correlators  are
\begin{align}
G^0_{\theta}(\tau_1-\tau_2) &= \langle \mathcal{T}  e^{i \left[\theta(\tau_1) -\theta(\tau_2) \right]}  \rangle_0  \notag\\
&= 
e^{N_0E_C (\tau_1-\tau_2)} e^{-E_C|\tau_1-\tau_2|/2},\label{eq:gtheta2}
\end{align}
Notice that, since the partition function of the free rotor $Z(N_0) = \sum_{N=-\infty}^{+\infty} e^{-E_C (N-N_0)^2/2}$ fulfills  $Z(N_0) = Z(N_0 \pm 1)$, we can restrict
$-\tfrac{1}{2} < N_0 < \tfrac{1}{2}$, which is necessary to ensure that the correlators decay to zero for $|\tau_1-\tau_2|\to +\infty$. Similarly, using the generating functional the four-point correlator can be obtained:

\begin{multline}
 \langle \mathcal{T}  e^{i \left[\theta(\tau_1) + \theta(\tau_2)  -\theta(\tau_3) -\theta(\tau_4) \right]}  \rangle_0  = 
 e^{N_0E_C \left(\tau_1 + \tau_2 - \tau_3 - \tau_4 \right)} \\ 
 \times e^{E_C\left(  |\tau_1-\tau_2| -   | \tau_1 - \tau_3| - |\tau_1  - \tau_4| - |\tau_2 - \tau_3| - |\tau_2-\tau_4| + |\tau_3 - \tau_4|\right)/2}. \label{eq:g4}
\end{multline}
These results are used in Appendix~\ref{app:pertchi}
in the perturbative calculation of the compressibility. 

\bibliography{references}

@article{Cazalilla_PhysRevLett.97.076401,
  title = {Dissipation-Driven Quantum Phase Transitions in a Tomonaga-Luttinger Liquid Electrostatically Coupled to a Metallic Gate},
  author = {Cazalilla, M. A. and Sols, F. and Guinea, F.},
  journal = {Phys. Rev. Lett.},
  volume = {97},
  issue = {7},
  pages = {076401},
  numpages = {4},
  year = {2006},
  month = {Aug},
  publisher = {American Physical Society},
  doi = {10.1103/PhysRevLett.97.076401},
  url = {https://link.aps.org/doi/10.1103/PhysRevLett.97.076401}
}

@article{Werner_JPSP,
author = {Werner ,Philipp and Troyer ,Matthias and Sachdev ,Subir},
title = {Quantum Spin Chains with Site Dissipation},
journal = {Journal of the Physical Society of Japan},
volume = {74},
number = {Suppl},
pages = {67-70},
year = {2005},
doi = {10.1143/JPSJS.74S.67},
URL = {https://doi.org/10.1143/JPSJS.74S.67},
eprint = {https://doi.org/10.1143/JPSJS.74S.67}
}

@article{Spohn1999,
	Author = {Spohn, Herbert and Zwerger, Wilhelm},
	Da = {1999/03/01},
	Doi = {10.1023/A:1004595419419},
	Id = {Spohn1999},
	Isbn = {1572-9613},
	Journal = {Journal of Statistical Physics},
	Number = {5},
	Pages = {1037--1043},
	Title = {Decay of the Two-Point Function in One-Dimensional O(N) Spin Models with Long-Range Interactions},
	Ty = {JOUR},
	Url = {https://doi.org/10.1023/A:1004595419419},
	Volume = {94},
	Year = {1999}}

@article{Lobos_PhysRevB.86.035455,
  title = {Magnetic phases in the one-dimensional Kondo chain on a metallic surface},
  author = {Lobos, Alejandro M. and Cazalilla, Miguel A. and Chudzinski, Piotr},
  journal = {Phys. Rev. B},
  volume = {86},
  issue = {3},
  pages = {035455},
  numpages = {13},
  year = {2012},
  month = {Jul},
  publisher = {American Physical Society},
  doi = {10.1103/PhysRevB.86.035455},
  url = {https://link.aps.org/doi/10.1103/PhysRevB.86.035455}
}

@article{Lobos3_PhysRevB.84.024523,
  title = {Superconductor-to-insulator transition in linear arrays of Josephson junctions capacitively coupled to metallic films},
  author = {Lobos, Alejandro M. and Giamarchi, Thierry},
  journal = {Phys. Rev. B},
  volume = {84},
  issue = {2},
  pages = {024523},
  numpages = {10},
  year = {2011},
  month = {Jul},
  publisher = {American Physical Society},
  doi = {10.1103/PhysRevB.84.024523},
  url = {https://link.aps.org/doi/10.1103/PhysRevB.84.024523}
}

@article{Lobos1_PhysRevB.80.214515,
  title = {Dissipation-driven phase transitions in superconducting wires},
  author = {Lobos, Alejandro M. and Iucci, An\'{\i}bal and M\"uller, Markus and Giamarchi, Thierry},
  journal = {Phys. Rev. B},
  volume = {80},
  issue = {21},
  pages = {214515},
  numpages = {10},
  year = {2009},
  month = {Dec},
  publisher = {American Physical Society},
  doi = {10.1103/PhysRevB.80.214515},
  url = {https://link.aps.org/doi/10.1103/PhysRevB.80.214515}
}

@article{Refael_PhysRevLett.101.106402,
  title = {Dissipation-Driven Quantum Phase Transition in Superconductor-Graphene Systems},
  author = {Lutchyn, Roman M. and Galitski, Victor and Refael, Gil and Das Sarma, S.},
  journal = {Phys. Rev. Lett.},
  volume = {101},
  issue = {10},
  pages = {106402},
  numpages = {4},
  year = {2008},
  month = {Sep},
  publisher = {American Physical Society},
  doi = {10.1103/PhysRevLett.101.106402},
  url = {https://link.aps.org/doi/10.1103/PhysRevLett.101.106402}
}

@article{Rosso1_PhysRevB.107.165113,
  title = {Bath-induced phase transition in a Luttinger liquid},
  author = {Majumdar, Saptarshi and Foini, Laura and Giamarchi, Thierry and Rosso, Alberto},
  journal = {Phys. Rev. B},
  volume = {107},
  issue = {16},
  pages = {165113},
  numpages = {8},
  year = {2023},
  month = {Apr},
  publisher = {American Physical Society},
  doi = {10.1103/PhysRevB.107.165113},
  url = {https://link.aps.org/doi/10.1103/PhysRevB.107.165113}
}

@article{Hofstetter_PhysRevLett.78.3737,
  title = {Single-Electron Box and the Helicity Modulus of an Inverse Square XY Model},
  author = {Hofstetter, W. and Zwerger, W.},
  journal = {Phys. Rev. Lett.},
  volume = {78},
  issue = {19},
  pages = {3737--3740},
  numpages = {0},
  year = {1997},
  month = {May},
  publisher = {American Physical Society},
  doi = {10.1103/PhysRevLett.78.3737},
  url = {https://link.aps.org/doi/10.1103/PhysRevLett.78.3737}
}

@article{Pankov_PhysRevB.69.054426,
  title = {Non-Fermi-liquid behavior from two-dimensional antiferromagnetic fluctuations: A renormalization-group and large-$N$ analysis},
  author = {Pankov, Sergey and Florens, Serge and Georges, Antoine and Kotliar, Gabriel and Sachdev, Subir},
  journal = {Phys. Rev. B},
  volume = {69},
  issue = {5},
  pages = {054426},
  numpages = {12},
  year = {2004},
  month = {Feb},
  publisher = {American Physical Society},
  doi = {10.1103/PhysRevB.69.054426},
  url = {https://link.aps.org/doi/10.1103/PhysRevB.69.054426}
}

@book{Giamarchi_book,
    author = {Giamarchi, Thierry},
    title = "{Quantum Physics in One Dimension}",
    publisher = {Oxford University Press},
    year = {2003},
    month = {12},
    isbn = {9780198525004},
    doi = {10.1093/acprof:oso/9780198525004.001.0001},
    url = {https://doi.org/10.1093/acprof:oso/9780198525004.001.0001},
}

@article{Cazalilla_JphysB,
doi = {10.1088/0953-4075/37/7/051},
url = {https://dx.doi.org/10.1088/0953-4075/37/7/051},
year = {2004},
month = {mar},
publisher = {},
volume = {37},
number = {7},
pages = {S1},
author = {M A Cazalilla},
title = {Bosonizing one-dimensional cold atomic gases},
journal = {Journal of Physics B: Atomic, Molecular and Optical Physics}
}

@article{Cazalilla_RevModPhys.83.1405,
  title = {One dimensional bosons: From condensed matter systems to ultracold gases},
  author = {Cazalilla, M. A. and Citro, R. and Giamarchi, T. and Orignac, E. and Rigol, M.},
  journal = {Rev. Mod. Phys.},
  volume = {83},
  issue = {4},
  pages = {1405--1466},
  numpages = {0},
  year = {2011},
  month = {Dec},
  publisher = {American Physical Society},
  doi = {10.1103/RevModPhys.83.1405},
  url = {https://link.aps.org/doi/10.1103/RevModPhys.83.1405}
}

@book{GogolinTsvelik1999,
  title = {Bosonization and Strongly Correlated Systems},
  publisher = {Cambridge University Press},
  year = {1999},
  author = {A. O. Gogolin and A. A. Nersesyan and A. M. Tsvelik},
  address = {Cambridge}
}

@article{LeDoussal_PhysRevB.82.155127,
  title = {Aharonov-Bohm effect in the presence of dissipative environments},
  author = {Horovitz, Baruch and Le Doussal, Pierre},
  journal = {Phys. Rev. B},
  volume = {82},
  issue = {15},
  pages = {155127},
  numpages = {15},
  year = {2010},
  month = {Oct},
  publisher = {American Physical Society},
  doi = {10.1103/PhysRevB.82.155127},
  url = {https://link.aps.org/doi/10.1103/PhysRevB.82.155127}
}

@article{Pruisken_PhysRevB.81.085428,
  title = {Macroscopic charge quantization in single-electron devices},
  author = {Burmistrov, I. S. and Pruisken, A. M. M.},
  journal = {Phys. Rev. B},
  volume = {81},
  issue = {8},
  pages = {085428},
  numpages = {21},
  year = {2010},
  month = {Feb},
  publisher = {American Physical Society},
  doi = {10.1103/PhysRevB.81.085428},
  url = {https://link.aps.org/doi/10.1103/PhysRevB.81.085428}
}

@article{Matveev1991,
  title = {Quantum fluctuations of the charge of a metal particle under the Coulomb blockade conditions},
  author = {Matveev, K. A.},
  journal = {Sov. Phys. Journal of Experimental and Theoretical Physics},
  volume = {72},
  issue = {5},
  pages = {892},
  numpages = {7},
  year = {1991},
  month = {Feb},
  publisher = {},
  doi = {},
  url = {http://www.jetp.ras.ru/cgi-bin/dn/e_072_05_0892.pdf}
}

@article{Renn2,
  title = {Quantum phase transitions in dissipative tunnel junctions},
  author = {Drewes, Scott and Arovas, Daniel P. and Renn, Scot},
  journal = {Phys. Rev. B},
  volume = {68},
  issue = {16},
  pages = {165345},
  numpages = {9},
  year = {2003},
  month = {Oct},
  publisher = {American Physical Society},
  doi = {10.1103/PhysRevB.68.165345},
  url = {https://link.aps.org/doi/10.1103/PhysRevB.68.165345}
}

@misc{Renn1,
      title={Large N behavior of Non-Ohmic Tunnel Junctions}, 
      author={S. R. Renn},
      year={1997},
      eprint={cond-mat/9708194},
      archivePrefix={arXiv},
      primaryClass={cond-mat}
}

@article{Nersesyan1978,
  author = {Dzhaparidze, G. I. and Nersesyan, A. A.},
  title = {Magnetic-field phase transition in a one-dimensional system of electrons
	with attraction},
  journal = {JETP Lett.},
  year = {1978},
  volume = {27},
  pages = {334},
  url = {http://www.jetpletters.ac.ru/ps/1549/article_23715.shtml}
}

@article{Pokrovsky1979,
  author = {Pokrovsky, V. L. and Talapov, A. L.},
  title = {Ground State, Spectrum, and Phase Diagram of Two-Dimensional Incommensurate
	Crystals},
  journal = {Phys. Rev. Lett.},
  year = {1979},
  volume = {42},
  pages = {65--67},
  number = {1},
  month = {Jan},
  doi = {10.1103/PhysRevLett.42.65},
  numpages = {2},
  publisher = {American Physical Society}
}

@article{Malatsetxebarria_PhysRevA.88.063630,
  title = {Dissipative effects on the superfluid-to-insulator transition in mixed-dimensional optical lattices},
  author = {Malatsetxebarria, E. and Cai, Zi and Schollw\"ock, U. and Cazalilla, Miguel A.},
  journal = {Phys. Rev. A},
  volume = {88},
  issue = {6},
  pages = {063630},
  numpages = {16},
  year = {2013},
  month = {Dec},
  publisher = {American Physical Society},
  doi = {10.1103/PhysRevA.88.063630},
  url = {https://link.aps.org/doi/10.1103/PhysRevA.88.063630}
}

@misc{ribeiro2023dissipationinduced,
      title={Dissipation-induced long-range order in the one-dimensional Bose-Hubbard model}, 
      author={Afonso L. S. Ribeiro and Paul McClarty and Pedro Ribeiro and Manuel Weber},
      year={2023},
      eprint={2311.07683},
      archivePrefix={arXiv},
      primaryClass={cond-mat.str-el}
}

@article{Sachdev_PhysRevLett.92.237003,
  title = {Universal Conductance of Nanowires near the Superconductor-Metal Quantum Transition},
  author = {Sachdev, Subir and Werner, Philipp and Troyer, Matthias},
  journal = {Phys. Rev. Lett.},
  volume = {92},
  issue = {23},
  pages = {237003},
  numpages = {4},
  year = {2004},
  month = {Jun},
  publisher = {American Physical Society},
  doi = {10.1103/PhysRevLett.92.237003},
  url = {https://link.aps.org/doi/10.1103/PhysRevLett.92.237003}
}

@article{Tucker_PhysRevB.3.3768,
  title = {Onset of Superconductivity in One-Dimensional Systems},
  author = {Tucker, J. R. and Halperin, B. I.},
  journal = {Phys. Rev. B},
  volume = {3},
  issue = {11},
  pages = {3768--3782},
  numpages = {0},
  year = {1971},
  month = {Jun},
  publisher = {American Physical Society},
  doi = {10.1103/PhysRevB.3.3768},
  url = {https://link.aps.org/doi/10.1103/PhysRevB.3.3768}
}

@article{Larkin_PhysRevLett.86.1869,
  title = {Quantum Superconductor-metal Transition in a Proximity Array},
  author = {Feigel'man, M. V. and Larkin, A. I. and Skvortsov, M. A.},
  journal = {Phys. Rev. Lett.},
  volume = {86},
  issue = {9},
  pages = {1869--1872},
  numpages = {0},
  year = {2001},
  month = {Feb},
  publisher = {American Physical Society},
  doi = {10.1103/PhysRevLett.86.1869},
  url = {https://link.aps.org/doi/10.1103/PhysRevLett.86.1869}
}

@article{FEIGELMAN1998107,
title = {Quantum  superconductor-metal transition in a 2D proximity-coupled array},
journal = {Chemical Physics},
volume = {235},
number = {1},
pages = {107-114},
year = {1998},
issn = {0301-0104},
doi = {https://doi.org/10.1016/S0301-0104(98)00075-5},
url = {https://www.sciencedirect.com/science/article/pii/S0301010498000755},
author = {M.V. Feigel'man and A.I. Larkin}
}

@article{Luther_PhysRevLett.33.589,
  title = {Backward Scattering in the One-Dimensional Electron Gas},
  author = {Luther, A. and Emery, V. J.},
  journal = {Phys. Rev. Lett.},
  volume = {33},
  issue = {10},
  pages = {589--592},
  numpages = {0},
  year = {1974},
  month = {Sep},
  publisher = {American Physical Society},
  doi = {10.1103/PhysRevLett.33.589},
  url = {https://link.aps.org/doi/10.1103/PhysRevLett.33.589}
}

@article{Cazalilla_2006,
doi = {10.1088/1367-2630/8/8/158},
url = {https://dx.doi.org/10.1088/1367-2630/8/8/158},
year = {2006},
month = {aug},
publisher = {},
volume = {8},
number = {8},
pages = {158},
author = {M A Cazalilla and A F Ho and T Giamarchi},
title = {Interacting Bose gases in quasi-one-dimensional optical lattices},
journal = {New Journal of Physics}
}

@article{Ho_PhysRevLett.92.130405,
  title = {Deconfinement in a 2D Optical Lattice of Coupled 1D Boson Systems},
  author = {Ho, A. F. and Cazalilla, M. A. and Giamarchi, T.},
  journal = {Phys. Rev. Lett.},
  volume = {92},
  issue = {13},
  pages = {130405},
  numpages = {4},
  year = {2004},
  month = {Apr},
  publisher = {American Physical Society},
  doi = {10.1103/PhysRevLett.92.130405},
  url = {https://link.aps.org/doi/10.1103/PhysRevLett.92.130405}
}

@article{Huang_PhysRevResearch.5.043192,
  title = {Modeling particle loss in open systems using Keldysh path integral and second order cumulant expansion},
  author = {Huang, Chen-How and Giamarchi, Thierry and Cazalilla, Miguel A.},
  journal = {Phys. Rev. Res.},
  volume = {5},
  issue = {4},
  pages = {043192},
  numpages = {14},
  year = {2023},
  month = {Dec},
  publisher = {American Physical Society},
  doi = {10.1103/PhysRevResearch.5.043192},
  url = {https://link.aps.org/doi/10.1103/PhysRevResearch.5.043192}
}

@article{TomitaTakahashi2017,
author = {Takafumi Tomita  and Shuta Nakajima  and Ippei Danshita  and Yosuke Takasu  and Yoshiro Takahashi},
title = {Observation of the {M}ott insulator to superfluid crossover of a driven-dissipative {B}ose-{H}ubbard system},
journal = {Science Advances},
volume = {3},
number = {12},
pages = {e1701513},
year = {2017},
doi = {10.1126/sciadv.1701513},
URL = {https://www.science.org/doi/abs/10.1126/sciadv.1701513},
eprint = {https://www.science.org/doi/pdf/10.1126/sciadv.1701513}}

@article{Uchino2022,
  title = {Comparative study for two-terminal transport through a lossy one-dimensional quantum wire},
  author = {Uchino, Shun},
  journal = {Phys. Rev. A},
  volume = {106},
  issue = {5},
  pages = {053320},
  numpages = {14},
  year = {2022},
  month = {Nov},
  publisher = {American Physical Society},
  doi = {10.1103/PhysRevA.106.053320},
  url = {https://link.aps.org/doi/10.1103/PhysRevA.106.053320}
}

@Article{SyassenDuerr2008,
  author = 	 {N. Syassen and D. M. Bauer and M. Lettner and T. Volz and D. Dietze and J. J. Garcia-Ripoll and J. I. Cirac and G. Rempe and S. D\"urr},
  title = 	 {Strong dissipation inhibits losses and induces correlations in cold molecular gases },
  journal = 	 {Science},
  year = 	 {2008},
  OPTkey = 	 {},
  volume = 	 {320},
  OPTnumber = 	 {},
  pages = 	 {1329},
  OPTmonth = 	 {},
  OPTnote = 	 {},
  OPTannote = 	 {}
}

@article{Guoetal2024,
  author  = {Guo, Yanliang and Yao, Hepeng and Ramanjanappa, Satwik and Dhar, Sudipta and Horvath, Milena and Pizzino, Lorenzo and Giamarchi, Thierry and Landini, Manuele and Nägerl, Hanns-Christoph},
  title   = {Observation of the 2D--1D crossover in strongly interacting ultracold bosons},
  journal = {Nature Physics},
  year    = {2024},
  volume  = {20},
  number  = {6},
  pages   = {934--938},
  doi     = {10.1038/s41567-024-02459-3}
}

@article{Haldane_PhysRevLett.47.1840,
  title = {Effective Harmonic-Fluid Approach to Low-Energy Properties of One-Dimensional Quantum Fluids},
  author = {Haldane, F. D. M.},
  journal = {Phys. Rev. Lett.},
  volume = {47},
  issue = {25},
  pages = {1840--1843},
  numpages = {0},
  year = {1981},
  month = {Dec},
  publisher = {American Physical Society},
  doi = {10.1103/PhysRevLett.47.1840},
  url = {https://link.aps.org/doi/10.1103/PhysRevLett.47.1840}
}

@article{Lukyanov_2006,
doi = {10.1088/1742-5468/2006/11/P11002},
url = {https://doi.org/10.1088/1742-5468/2006/11/P11002},
year = {2006},
month = {nov},
publisher = {},
volume = {2006},
number = {11},
pages = {P11002},
author = {Lukyanov, Sergei L and Werner, Philipp},
title = {Universal scaling behaviour of the single electron box in the strong tunnelling
limit},
journal = {Journal of Statistical Mechanics: Theory and Experiment}
}

@article{Grabert_PhysRevLett.81.2324,
  title = {Effect of Tunneling Conductance on the Coulomb Staircase},
  author = {G\"oppert, Georg and Grabert, Hermann and Prokof'ev, Nikolai V. and Svistunov, Boris V.},
  journal = {Phys. Rev. Lett.},
  volume = {81},
  issue = {11},
  pages = {2324--2327},
  numpages = {0},
  year = {1998},
  month = {Sep},
  publisher = {American Physical Society},
  doi = {10.1103/PhysRevLett.81.2324},
  url = {https://link.aps.org/doi/10.1103/PhysRevLett.81.2324}
}

@article{Dupuis2026,
  title = {Functional renormalization group study of a dissipative Bose-Hubbard model},
  author = {Bouverot-Dupuis, Oscar and Grison, Vincent and Paris, Nicolas},
  journal = {Phys. Rev. B},
  volume = {113},
  issue = {17},
  pages = {174503},
  numpages = {14},
  year = {2026},
  month = {May},
  publisher = {American Physical Society},
  doi = {10.1103/9t2w-8zzd},
  url = {https://link.aps.org/doi/10.1103/9t2w-8zzd}
}

@article{Fisher_PhysRevB.40.546,
  title = {Boson localization and the superfluid-insulator transition},
  author = {Fisher, Matthew P. A. and Weichman, Peter B. and Grinstein, G. and Fisher, Daniel S.},
  journal = {Phys. Rev. B},
  volume = {40},
  issue = {1},
  pages = {546--570},
  numpages = {0},
  year = {1989},
  month = {Jul},
  publisher = {American Physical Society},
  doi = {10.1103/PhysRevB.40.546},
  url = {https://link.aps.org/doi/10.1103/PhysRevB.40.546}
}
\end{document}